\documentclass[apjl]{emulateapj}
\usepackage{amsfonts,amsmath,graphicx,natbib,apjfonts,subfigure}

\def\ra#1#2#3{#1$^{\rm h}$#2$^{\rm m}$#3$^{\rm s}$}
\def\dec#1#2#3{#1$^\circ$#2$'$#3$''$}

\def\swift{{\it Swift}}
\def\chandra{{\it Chandra}}
\def\grb{GRB\,111117A}

\begin{document}
\title{The Afterglow and Environment of the Short GRB\,111117A}
\author{
R.~Margutti\altaffilmark{1},
E.~Berger\altaffilmark{1},
W.~Fong\altaffilmark{1}, 
B.~A.~Zauderer\altaffilmark{1},
S.~B.~Cenko\altaffilmark{2},
J.~Greiner\altaffilmark{3},
A.~M.~Soderberg\altaffilmark{1},
A.~Cucchiara\altaffilmark{4},
A.~Rossi\altaffilmark{5}, 
S.~Klose\altaffilmark{5},
S.~Schmidl\altaffilmark{5}, 
D.~Milisavljevic\altaffilmark{1},
\& N.~Sanders\altaffilmark{1}
}

\altaffiltext{1}{Harvard-Smithsonian Center for Astrophysics, 60
Garden Street, Cambridge, MA 02138, USA}

\altaffiltext{2}{Department of Astronomy, University of California,
Berkeley, CA 94720, USA}

\altaffiltext{3}{Max-Planck-Institut f\"ur Extraterrestrische Physik, 85740 Garching,
Germany}

\altaffiltext{4}{Department of Astronomy and Astrophysics, UCO/Lick
Observatory, University of California, 1156 High Street, Santa Cruz,
CA 95064, USA}

\altaffiltext{5}{Th\"uringer Landessternwarte Tautenburg, Sternwarte 5,
07778 Tautenburg, Germany}

\begin{abstract} We present multi-wavelength observations of the
afterglow of the short GRB\,111117A, and follow-up observations of its
host galaxy.  From rapid optical and radio observations we place
limits of $r\gtrsim 25.5$ mag at $\delta t\approx 0.55$ d and
$F_\nu(5.8\,{\rm GHz}) \lesssim 18$ $\mu$Jy at $\delta t\approx 0.50$
d, respectively.  However, using a {\it Chandra} observation at
$\delta t\approx 3.0$ d we locate the absolute position of the X-ray
afterglow to an accuracy of $0.22''$ ($1\sigma$), a factor of about
$6$ times better than the \swift/XRT position.  This allows us to
robustly identify the host galaxy and to locate the burst at a
projected offset of $1.25\pm 0.20''$ from the host centroid.  Using
optical and near-IR observations of the host galaxy we determine a
photometric redshift of $z=1.3^{+0.3}_{-0.2}$, one of the highest for
any short GRB, and leading to a projected physical offset for the
burst of $10.5\pm 1.7$ kpc, typical of previous short GRBs.  At this
redshift, the isotropic $\gamma$-ray energy is $E_{\rm\gamma,
iso}\approx 3.0\times 10^{51}$ erg (rest-frame $23-2300$ keV) with a
peak energy of $E_{\rm pk}\approx 850-2300$ keV (rest-frame).  In
conjunction with the isotropic X-ray energy, \grb\ appears to follow
our recently-reported $E_{\rm x,iso}$-$E_{\rm
\gamma,iso}$-$E_{\rm{pk}}$ universal scaling.  Using the X-ray data
along with the optical and radio non-detections we find that for a
blastwave kinetic energy of $E_{\rm K,iso}\approx E_{\rm\gamma,iso}$
erg, the circumburst density is $n_0\approx 3\times 10^{-4}-1$
cm$^{-3}$ (for a range of $\epsilon_B=0.001-0.1$).  Similarly, from
the non-detection of a break in the X-ray light curve at $\delta
t\lesssim 3$ d, we infer a minimum opening angle for the outflow of
$\theta_j\gtrsim 3-10^\circ$ (depending on the circumburst density).
We conclude that \emph{Chandra} observations of short GRBs are
effective at determining precise positions and robust host galaxy
associations in the absence of optical and radio detections.
\end{abstract} 

\section{Introduction}
\label{Sec:Intro}

Precise localizations of short-duration gamma-ray bursts (GRB) are
critical for studies of their explosion properties, environments, and
progenitors.  In particular, such localizations can provide secure
associations with host galaxies, and hence redshift and offset
measurements (e.g., \citealt{Berger07,Fong10,Berger11b}).  To date,
most sub-arcsecond positions for short GRBs have relied on the
detection of optical afterglows (e.g.,
\citealt{Berger05,Hjorth05,Soderberg06}).  However, X-ray emission is
detected from a larger fraction of short bursts, and therefore
observations with the \chandra\ X-ray Observatory can equally provide
precise positions even in the absence of optical detections.  Indeed,
\chandra\ detections have been previously made for short GRBs 050709,
050724, 051221A, 080503, and 111020A
\citep{Fox05,Berger05,Burrows06,Grupe06,Soderberg06,Perley09,Fong12}, but only
in the latter case \chandra\ provided the sole route to a precise
position \citep{Fong12}.  In the other 4 cases, the afterglow was also
detected in the optical, as well as in the radio for GRBs 050724 and
051221A \citep{Berger05,Soderberg06}.

The  advantage of precise X-ray positions is that the X-ray
flux is potentially independent of the circumburst density if the
synchrotron cooling frequency is located redward of the X-ray band.
Thus, X-ray detections can in principle reduce any bias for small
offsets that may arise from optical detections, which do depend on the
density (although, see \citealt{Berger10} for short GRBs with optical
afterglows and evidence for large offsets of $\sim 50-100$ kpc).

Here, we present a \chandra\ detection of the X-ray afterglow of short
GRB\,111117A at $\delta t\approx 3$ d, which leads to a robust 
association with a galaxy at a
photometric redshift of $z\approx 1.3$ and to a precise offset
measurement.  Among short GRBs, only GRBs 050724 and 
051221A were detected at later times in X-rays.
Using the \chandra\ and \swift/XRT data we study the
properties of the X-ray afterglow in the context of the short GRB
sample, and in conjunction with deep optical and radio upper limits we
place constraints on the circumburst density.  Similarly,
optical/near-IR observations of the host allow us to determine its
physical properties (star formation rate, stellar mass, stellar
population age).

Throughout the paper we use the convention $F_{\nu}(\nu,t)\propto
\nu^{\beta}\,t^{\alpha}$, where the spectral energy index is related
to the spectral photon index by $\Gamma=1-\beta$.  All uncertainties
are quoted at 68\% confidence level, unless otherwise noted.
Magnitudes are reported in the AB system and have been corrected for
Galactic extinction \citep{Schlafly11}.  Finally, we use the standard
cosmological parameters: $H_{0}=71$ km s$^{-1}$ Mpc$^{-1}$,
$\Omega_{\Lambda}=0.73$, and $\Omega_{\rm M}=0.27$.


\section{Observations and Analysis}
\label{Sec:Obs}

\begin{figure}
\vskip -0.0 true cm
\centering
\includegraphics[scale=0.5]{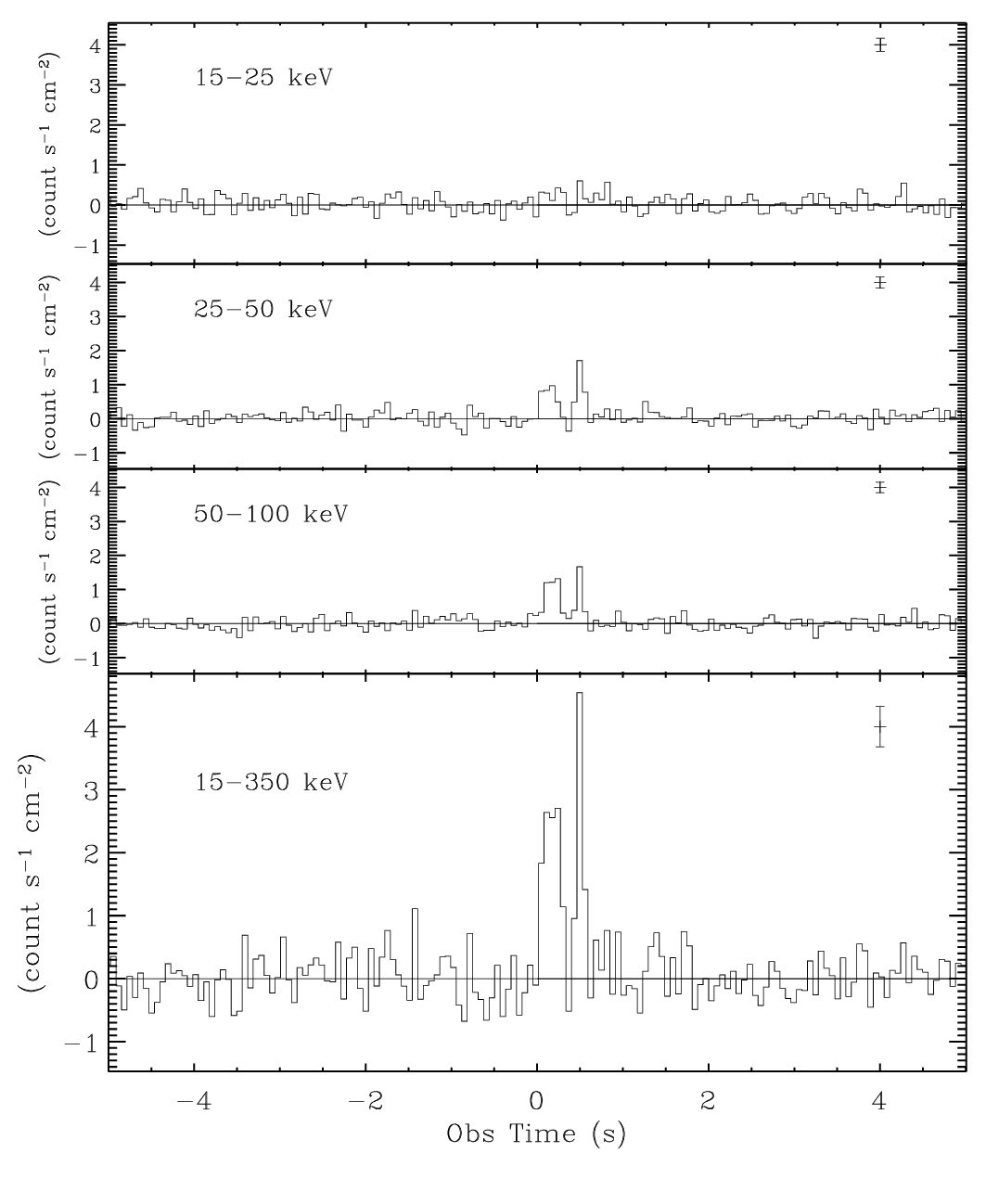}
\caption{\swift/BAT mask weighted light-curve in different energy
bands (binning time of 64 ms).  The typical $1\sigma$ error-bar is
shown in each panel.}  
\label{Fig:BATlc}
\end{figure}

\begin{figure*}
\vskip -0.0 true cm
\centering
\includegraphics[scale=0.3]{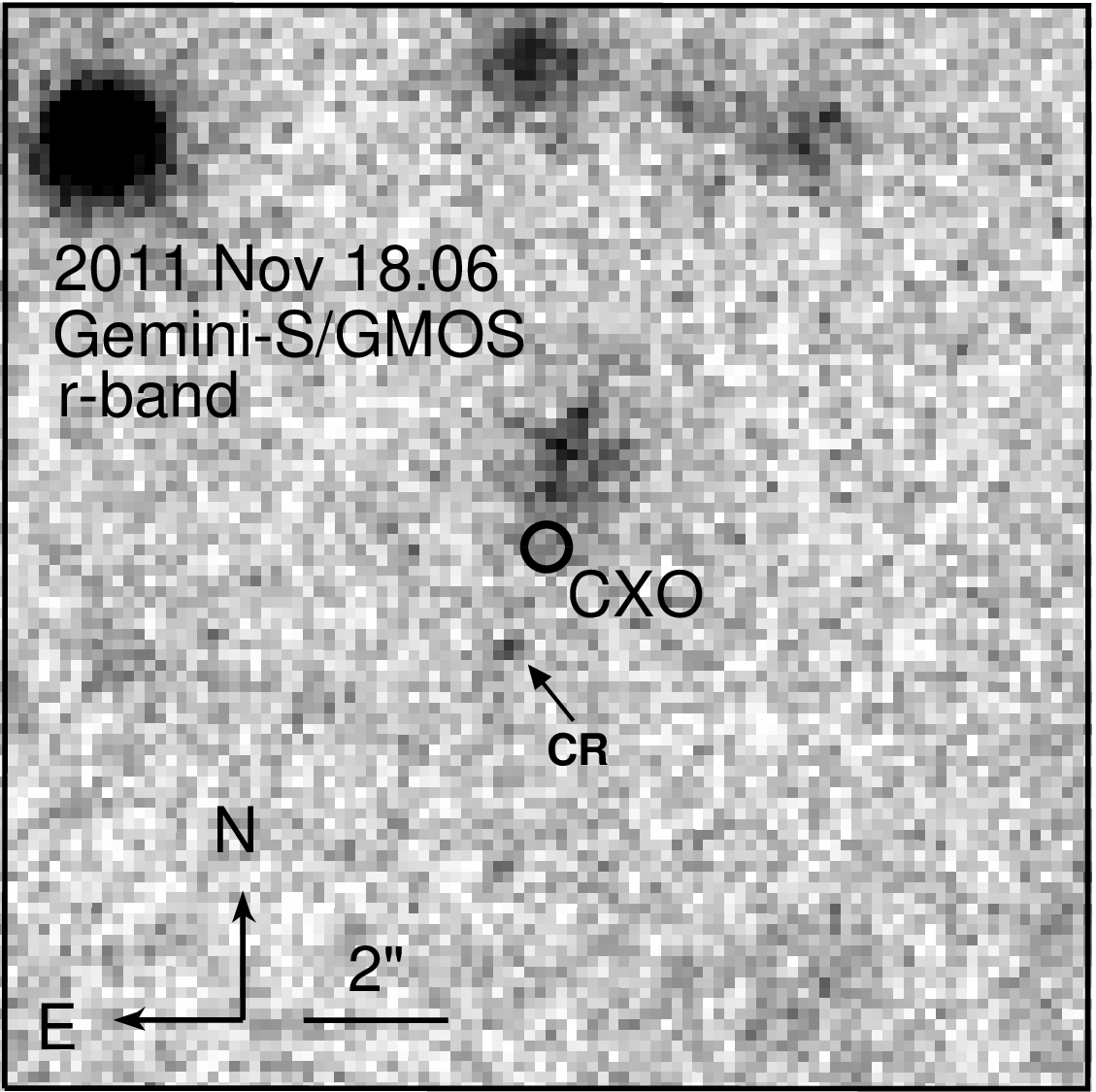}
\includegraphics[scale=0.3]{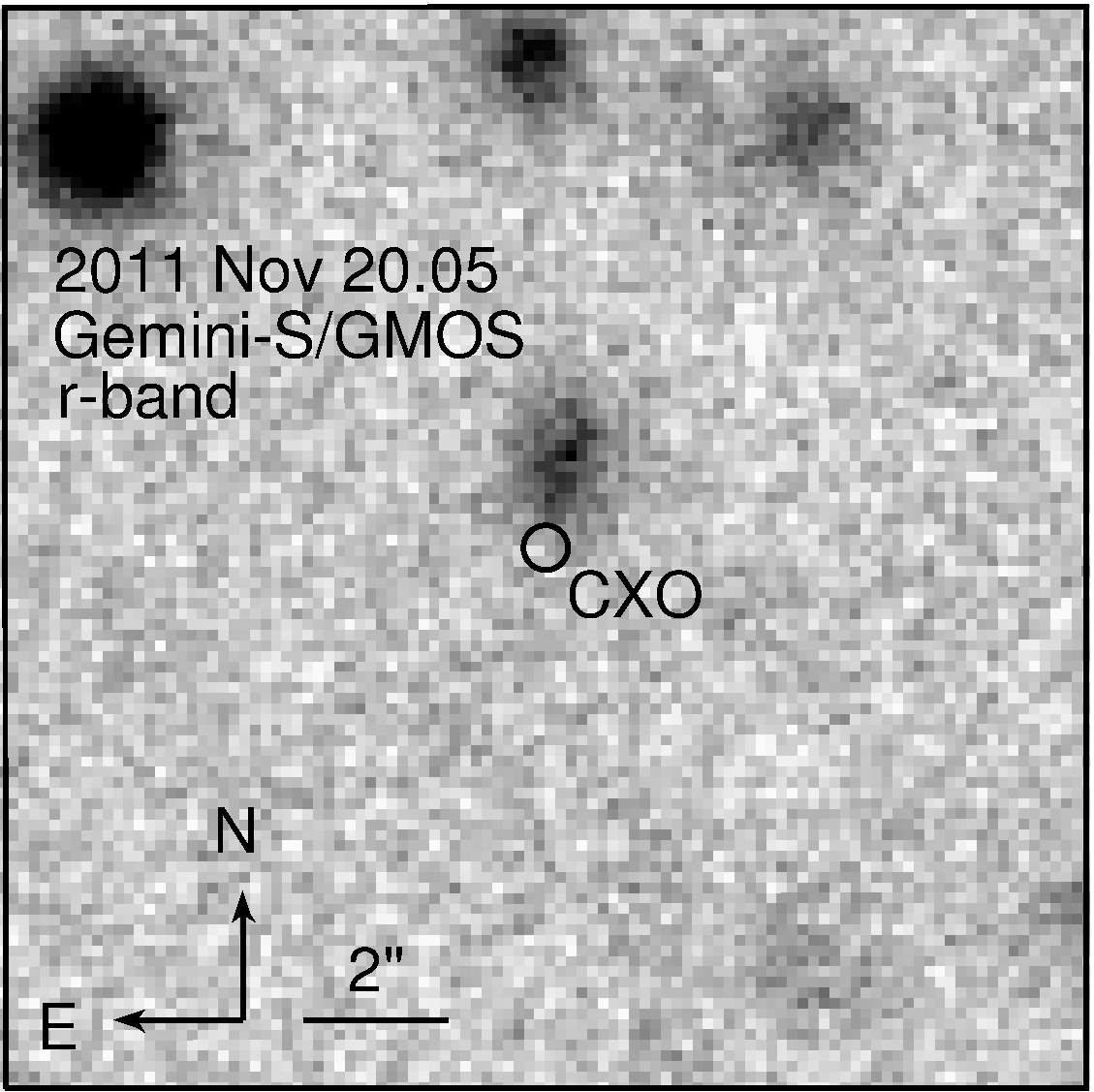}
\includegraphics[scale=0.3]{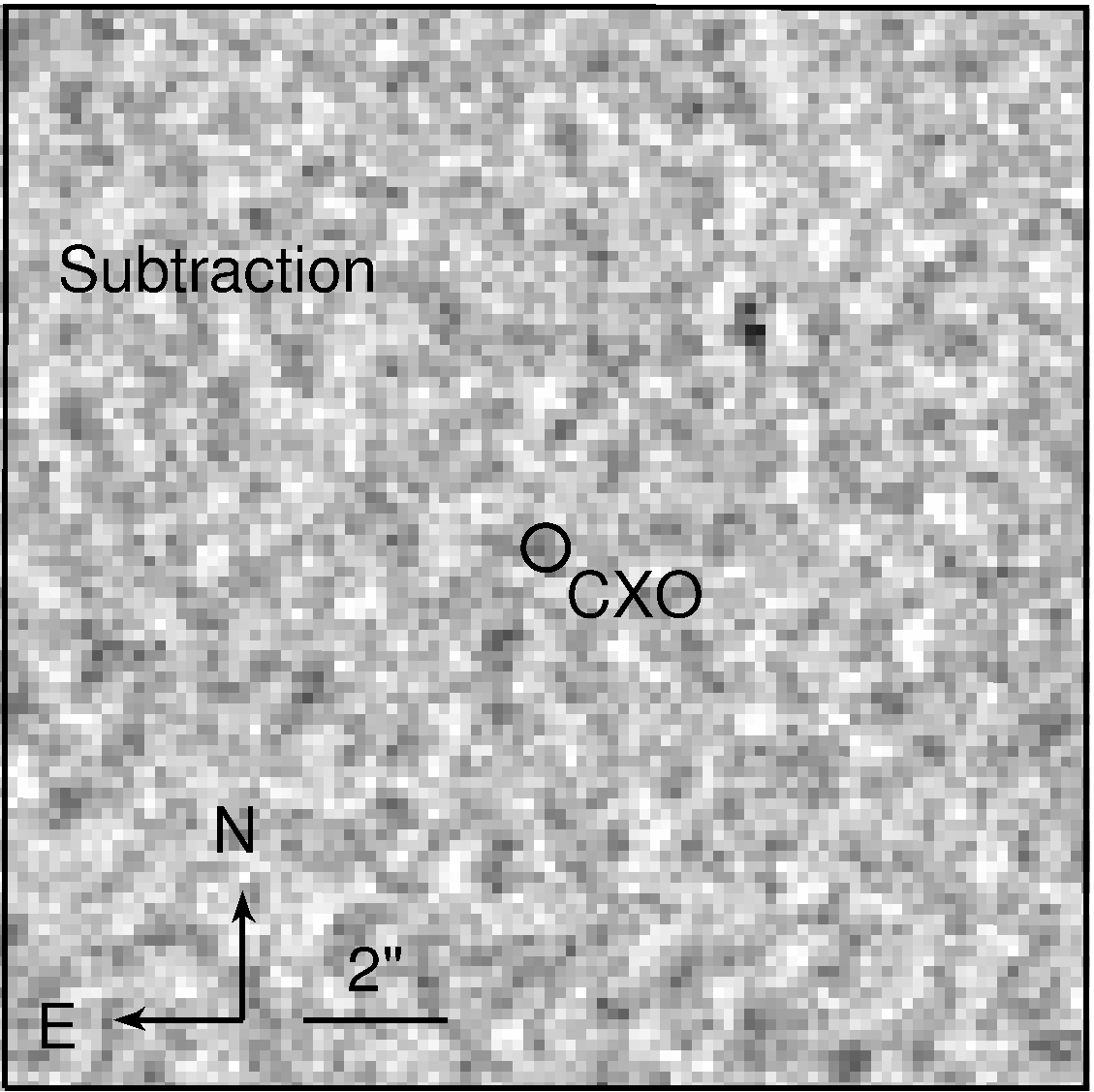}
\caption{Gemini-South $r$-band observations of \grb\ at $\delta
t\approx 0.55$ d (left) and $\approx 2.54$ d (middle).  Digital image
subtraction of the two observations (right) reveals no fading source
coincident with the \chandra\ position (circle) or the host galaxy.
We note that the apparent emission south-east of the \chandra\
position in the first epoch is a cosmic-ray (CR).}
\label{Fig:subtraction}
\end{figure*}

GRB\,111117A was detected on 2011 November 17.510 UT \citep{Mangano11}
by the Burst Alert Telescope (BAT; \citealt{Barthelmy05}) on-board the
\swift\ satellite \citep{Gehrels04}, with a ground-calculated
positional accuracy of $1.7'$ radius ($90\%$ containment;
\citealt{Sakamoto11}).  The burst was also detected by the
\emph{Fermi} Gamma-Ray Burst Monitor (GBM) in the energy range 10-1000
keV \citep{Foley11}. 

Follow-up observations with the X-ray Telescope (XRT;
\citealt{Burrows05}) commenced at $\delta t\approx 80$ s and
resulted in the detection of a fading X-ray source located at
RA=\ra{00}{50}{46.22} and Dec=$+$\dec{23}{00}{39.2} (J2000), with an
uncertainty of $2.1''$ radius ($90\%$ containment;
\citealt{Melandri11}).GRB111117A is therefore part of the handful of short GRBs
promptly re-pointed by \swift/XRT  and with broad band spectral coverage during the
prompt emission.
 The UV-Optical Telescope (UVOT;
\citealt{Roming05}) began observations of the field at $\delta
t\approx 137$ s but no corresponding source was found within the XRT
position to limits of $\gtrsim 19.5-21.5$ mag in the first 1300 s
\citep{Oates11}.

Ground-based optical observations commenced at $\delta t\approx 2$ hr
with a non-detection at $R\approx 22.2$ mag \citep{Zhao11}, and
eventually led to the detection of an extended source with $r\approx
24$ mag, identified as the potential host galaxy of GRB\,111117A
\citep{Andersen11,Fong11,Cenko11,Schmidl11,Melandri11b}.

Finally, a \chandra\ observation at $\delta t\approx 3.0$ d led to the
detection of the X-ray afterglow \citep{Sakamoto11b}, and a refined
analysis relative to our optical images from the Magellan 6.5-m
telescope provided a correction to the native \emph{Chandra}
astrometry and an initial offset from the host galaxy of about $1''$
\citep{Berger11}.  The analysis presented here supercedes the
\emph{Chandra} position quoted in the GCN circulars \citep{Berger11}.

\subsection{$\gamma$-ray Observations}
\label{SubSec:BAT}

We processed the \swift/BAT data with the latest version of the
HEASOFT package (v6.11), using the {\tt batgrbproduct} script to
generate event lists and quality maps for the 64 ms mask-weighted and
background-subtracted light curves (Figure~\ref{Fig:BATlc}).  The
ground-refined coordinates provided by \citet{Sakamoto11} were
adopted, and standard filtering and screening criteria were applied.
We also used the mask-weighting procedure to produce weighted,
background-subtracted spectra.

We find that the $\gamma$-ray emission consists of two pulses with a
total duration of $T_{90}=0.47\pm 0.05$ s ($15-350$ keV;
Figure~\ref{Fig:BATlc}), classifying GRB\,111117A as a short burst.
The spectral time-lag between the $100-350$ and $25-50$ keV bands is
$(0.6 \pm 2.4)$ ms, typical of short GRBs \citep{Sakamoto11}.  The
time-averaged spectrum in the $15-150$ keV range is best fit by a
single power-law model with a hard power-law index,
$\Gamma_{\gamma}=0.59\pm 0.14$.  The $\gamma$-ray fluence derived from
this spectrum is $F_\gamma=(1.3\pm 0.2)\times 10^{-7}$ erg cm$^{-2}$
in the $15-150$ keV band, in agreement with the values reported by
\cite{Sakamoto11} and \cite{Mangano11b}.

The {\it Fermi}/GBM spectrum in the energy range $10-1000$ keV is best
fit by a power-law with an exponential cut-off, with
$\Gamma_{\gamma}=0.69\pm 0.16$ \citep{Foley11}; this is consistent
with the BAT spectrum.  The peak energy is $E_{\rm{pk}}\gtrsim 370$
keV, while the observed exponential cut-off indicates that
$E_{\rm{pk}}\lesssim 1$ MeV.  The $10-1000$ keV band fluence derived
from this spectrum is $F_\gamma=(6.7\pm 0.2)\times 10^{-7}$ erg
cm$^{-2}$. We refer to \cite{Sakamoto12} for a detailed analysis
of the {\it Fermi}/GBM data.

\subsection{X-ray Observations}
\label{SubSec:XObs}

We analyzed the data using HEASOFT (v6.11) with standard filtering and
screening criteria, and generated the $0.3-10$ keV count-rate light
curve following the procedures in \cite{Margutti12}.  Our re-binning
scheme ensures a minimum signal-to-noise ratio $S/N=3$ for each
temporal bin.  The low count statistics of the Windowed Timing (WT)
observations do not allow us to constrain the spectral parameters
during the interval $\delta t\approx 80-87$ s.  We model the
time-averaged spectrum in the interval $87$ s to $40$ ks (total
exposure of about $9$ ks in the Photon Counting mode\footnote{See
\cite{Hill04} for \swift/XRT observing modes.}) with an absorbed
power-law model ($tbabs*ztbabs*pow$ within {\tt Xspec}) with a
best-fit spectral photon index of $\Gamma_{x}=2.0\pm 0.2 $ and an
intrinsic neutral hydrogen column density of $N_{\rm H,int}=(6.7\pm
3.0)\times 10^{21}$ cm$^{-2}$ ($C-stat=98$ for 152 d.o.f.) in excess
to the Galactic column density, $N_{\rm H,MW}=3.7\times 10^{20}$
cm$^{-2}$ \citep{Kalberla05}; we adopt the best-fit photometric
redshift of $z=1.3$ derived in \S\ref{Sec:Host}.  From the best-fit
spectrum we derive an unabsorbed count-to-flux conversion factor of
$6.5\times 10^{-11}$ erg cm$^{-2}$ counts$^{-1}$ (0.3-10 keV).
Uncertainties arising from the flux calibration procedure have been
propagated into the individual error-bars.
 
We analyzed the public \chandra\ data (PI: Sakamoto) with the {\tt
CIAO} software package (v4.3), using the calibration database CALDB
v4.4.2, and applying standard ACIS data filtering.  Using {\tt
wavdetect} we detect a source at $3.4\sigma$ significance at a
position consistent with the XRT afterglow, with a net count-rate of
$(3.3\pm 1.3)\times 10^{-4}$ counts s$^{-1}$ ($0.5-8$ keV; total exposure
time of $19.8$ ks).  Assuming the spectral parameters from the XRT
analysis, this translates to an unabsorbed flux of $(3.5\pm 1.4)\times
10^{-15}$ erg s$^{-1}$ cm$^{-2}$ ($0.3-10$ keV).

The X-ray light curve (Figure~\ref{Fig:XRTlc}) exhibits an overall
single power-law decay, with an apparent flare at $\delta t\approx
150$ s ($\approx 3\sigma$ confidence level).  The best-fit power-law
index at $\delta t\gtrsim 300$ s is $\alpha_x=-1.21\pm 0.05$
($\chi^2=7.6$ for $11$ d.o.f.)\footnotemark\footnotetext{The power-law
index is obtained by minimizing the integral of the model over the
effective duration of each temporal bin of the light-curve.  This
procedure is of primary importance in the case of bins with long
duration and produces more accurate results than the standard $\chi^2$
procedure, which compares the model and the data at the nominal bin
time, but does not consider the finite bin duration and the evolution
of the model during the time interval.}.

\subsection{Optical Afterglow Limits}
\label{sec:aglim}

We obtained deep $r$-band observations with the Gemini Multi-Object
Spectrograph (GMOS; \citealt{Hook04}) mounted on the Gemini-South 8-m
telescope on 2011 November 18.06 and 20.05 UT, with total exposure
times of 1800 s and 2880 s, respectively.  We processed the data using
the {\tt gemini} package in IRAF\footnote{IRAF is distributed by the National Optical
Astronomy Observatory, which is operated by the Association for Research
in Astronomy, Inc., under cooperative agreement with the National Science
Foundation.}, and calibrated the photometry with
several nearby point sources from the Sloan Digital Sky Survey (SDSS;
\citealt{Abazajian09}).  We further performed digital image
subtraction of the two epochs using the High Order Transformation and
Point Spread Function and Template Subtraction
(HOTPANTS\footnotemark\footnotetext{http://www.astro.washington.edu/users/becker/hotpants.html.}),
but no fading source is detected within the XRT error circle, or in
coincidence with the putative host galaxy to $r\gtrsim 25.5$ mag
($3\sigma$) at $\delta t\approx 0.55$ d
(Figure~\ref{Fig:subtraction}).  We note that this is the deepest
limit to date on the early optical emission from a short GRB
\citep{Berger10,Fong11b}, with the exception of GRB\,080503 which was
eventually detected at $\delta t\gtrsim 1$ d \citep{Perley09}.
Indeed, the median optical afterglow brightness for detected short
GRBs on a similar timescale is $r\approx 23.5$ mag, a factor of at
least 6 times brighter \citep{Berger10,Fong11b}.

\subsection{X-ray/Optical Differential Astrometry}
\label{SubSec:astrometry}

\begin{figure}
\vskip -0.0 true cm
\centering
\includegraphics[scale=0.4]{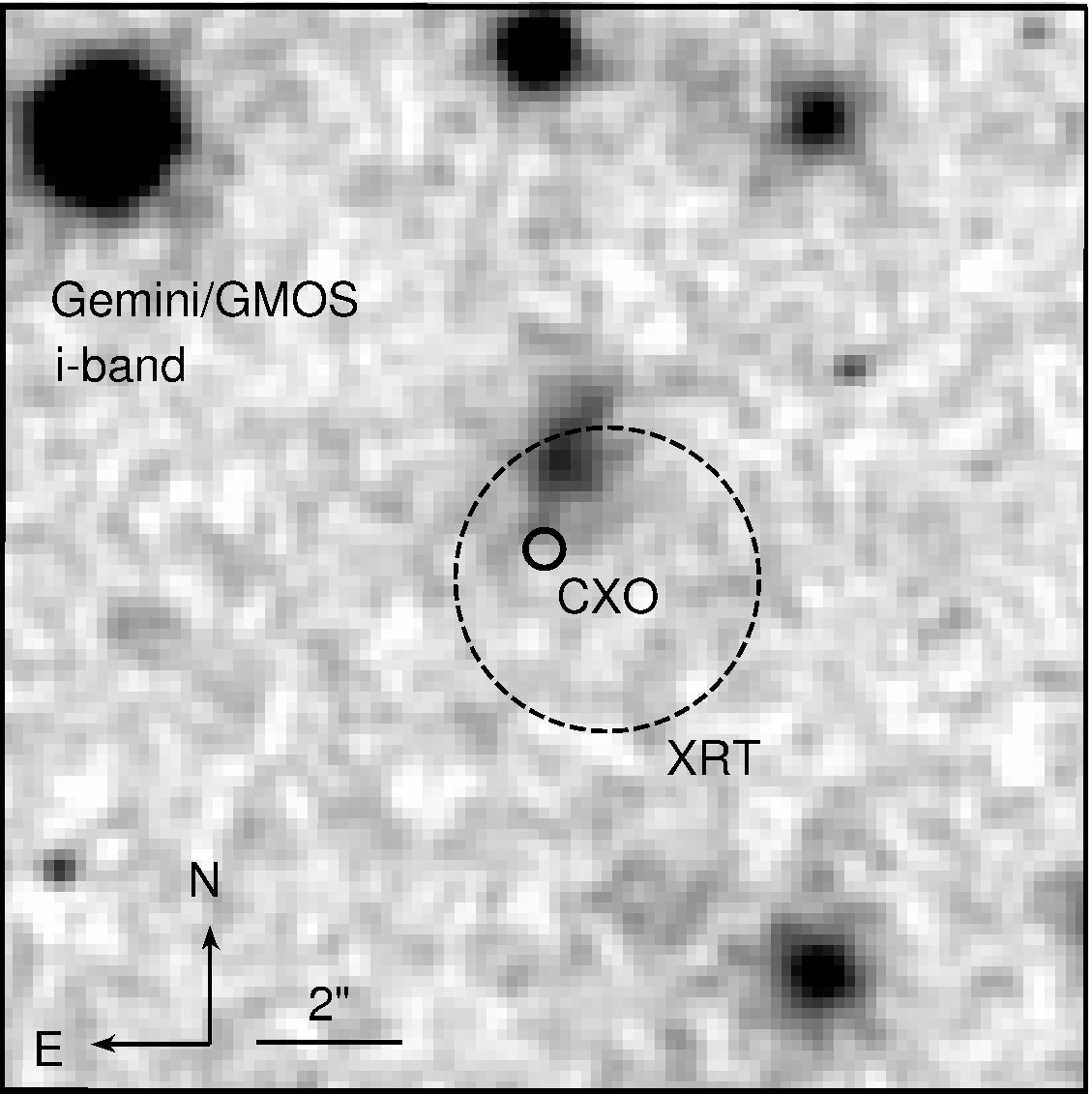}
\caption{Gemini/GMOS $i$-band image of a $15''\times 15''$ field
centered on the \chandra\ position of the X-ray afterglow of
GRB\,111117A (solid circle; $90\%$ confidence region).  Also shown is
the \swift/XRT position (dashed circle; $90\%$ confidence region).
The host galaxy is clearly detected with an offset of about $1.25''$
from the \chandra\ position.}
\label{Fig:image}
\end{figure}

In the absence of an optical afterglow we use the \chandra\
observation to refine the \swift/XRT position to sub-arcsecond
accuracy.  We perform differential astrometry between the \chandra\
data and a Gemini-North $i$-band observation (\S\ref{SubSec:OptObs})
to determine the relative positions of the afterglow and host galaxy,
as well as to refine the native \chandra\ astrometry.  We use {\tt
SExtractor}\footnotemark\footnotetext{http://sextractor.sourceforge.net/}
to determine the positions and centroid uncertainties of sources in
the GMOS image.  Performing an absolute astrometric tie to the Sloan
Digital Sky Survey (SDSS) catalog using 71 common point sources, we
find a resulting rms value of $\sigma_{\rm GMOS-SDSS}=0.13''$
($0.09''$ in each coordinate).

To refine the native \chandra\ astrometry and to determine the
location of the X-ray afterglow relative to the GMOS image, we perform
differential astrometry.  We use the {\tt CIAO} routine {\tt
wavdetect} to obtain positions and $1\sigma$ centroid uncertainties of
X-ray sources in the field, including the afterglow of GRB\,111117A,
with a resulting $\sigma_{{\rm X,ag}}=0.13''$.  We calculate an
astrometric tie based on two X-ray and optically bright common sources
and find weighted mean offsets of $\delta {\rm RA}=-0.18\pm 0.03''$
and $\delta {\rm Dec}=-0.02\pm 0.11''$ giving a tie
uncertainty\footnotemark\footnotetext{There are two additional fainter
common sources, which lead to increased scatter in the astrometric tie
without changing the absolute value, and we therefore use only the two
bright sources.} of $\sigma_{\rm CXO-GMOS}= 0.12''$.  Applying the
astrometric solution we obtain a \chandra\ afterglow position of
RA=\ra{00}{50}{46.283} and Dec=$+$\dec{23}{00}{39.64}; see
Figure~\ref{Fig:image}.  The total $1\sigma$ uncertainty in the
absolute position is $0.22''$, accounting for the SDSS-GMOS
astrometric tie, GMOS-\chandra\ tie, and the X-ray afterglow
positional uncertainty.  The \chandra\ position is consistent with the
XRT position, but refines its uncertainty by about a factor of 6.  We
note that the {\it relative} position of the X-ray afterglow in the
GMOS astrometric frame is $0.18''$ (GMOS-\chandra\ tie and afterglow
positional uncertainty only).

\subsection{Host Galaxy Optical/Near-IR Observations} 
\label{SubSec:OptObs}

\begin{deluxetable*}{lcccccccc}
\tabletypesize{\scriptsize}
\tablecolumns{9}
\tabcolsep0.0in\scriptsize
\tablewidth{0pc}
\tablecaption{GRB\,111117A Host Galaxy Optical and NIR Photometry
\label{tab:optnir}}
\tablehead {
\colhead {Date} &
\colhead {$\Delta t$}            &
\colhead {Telescope}             &
\colhead {Instrument}            &
\colhead {Filter}                &
\colhead {Exposures}             &
\colhead {$\theta_{\rm FWHM}$}   &
\colhead {Host$\,^a$}            &
\colhead {$A_{\lambda}$}        \\
\colhead {(UT)}                  &
\colhead {(d)}                   &
\colhead {}                      &
\colhead {}                      &
\colhead {}                      &
\colhead {(s)}                   &
\colhead {(arcsec)}              &
\colhead {(AB mag)}              &
\colhead {(AB mag)}  
}
\startdata
2011 Nov.~18.08 & $0.57$  & Magellan/Baade & IMACS    & $r$   & $4\times 300$  & $1.06$ & $23.72\pm 0.11$ & $0.069$ \\
2011 Nov.~28.13 & $10.62$ & MPG/ESO & GROND & $g$ & $8\times375$ & $1.07$ &$24.14\pm0.18$ & $0.120$\\
2011 Nov.~28.25 & $10.74$ & Gemini-North   & GMOS     & $i$   & $10\times 180$ & $0.75$ & $23.66\pm 0.08$ & $0.051$ \\
2011 Nov.~28.27 & $10.76$ & Gemini-North   & GMOS     & $z$   & $10\times 180$ & $0.83$ & $23.04\pm 0.18$ & $0.038$ \\
2011 Dec.~7.07  & $19.56$ & Magellan/Baade & FourStar & $J$   & $15\times 60$  & $0.78$ & $22.55\pm 0.30$ & $0.027$ \\
2011 Dec.~7.05  & $19.54$ & Magellan/Baade & FourStar & $K_s$ & $10\times 90$  & $0.68$ & $\gtrsim 22.10$ & $0.011$ 
\tablecomments{$^a$ These values have been corrected for Galactic
extinction, $A_\lambda$ \citep{Schlafly11}.}
\end{deluxetable*}

We obtained optical observations in the $griz$ bands, and near-IR
observations in $JK_s$ bands to determine the properties of the host
galaxy.  The details of the observations are summarized in
Table~\ref{tab:optnir}. The $g$-band observation was performed
with GROND (\citealt{Greiner08}) mounted at the 2.2 m MPG/ESO telescope at
La Silla Observatory (Chile). The $r$-band observation was obtained with
the Inamori-Magellan Areal Camera and Spectrograph (IMACS) on the
Magellan/Baade 6.5-m telescope, while the $iz$ band observations were
performed with GMOS mounted on the Gemini-North 8-m telescope.
Finally, the $JK_s$-band observations were obtained with the FourStar
wide-field near-IR camera on the Magellan/Baade telescope.  The GMOS
data were reduced using the {\tt gemini} package in IRAF, the IMACS
and GROND 
data were reduced using standard packages in IRAF, and the FourStar
data were reduced using a custom pipeline in python.

We identify a galaxy near the \chandra\ position, at
RA=\ra{00}{50}{46.267} and Dec=$+$\dec{23}{00}{40.87} (astrometry
relative to SDSS; \S\ref{SubSec:astrometry}), with a centroid
uncertainty of $0.08''$.  The offset between the galaxy centroid and
the \chandra\ afterglow position is $1.25''\pm 0.20''$.  Photometry of
the galaxy is performed in a $2''$ radius aperture, with the
zero-point determined by common sources with SDSS ($griz$) and 2MASS
($JK_s$). The resulting magnitudes are listed in
Table~\ref{tab:optnir}.

To determine the probability of chance coincidence for this galaxy
relative to the afterglow position we adopt the methodology of
\citet{Bloom02} and \citet{Berger10}.  The expected number density of
galaxies brighter than the apparent magnitude of the galaxy,
$m_r=23.6$ mag, is \citep{Hogg97,Beckwith06}:
\begin{equation}
\sigma(\le m)=\frac{1}{0.33\times {\rm ln}(10)}\times
10^{0.33(m_r-24)-2.44}\approx 0.004 \,\,\,\,{\rm
arcsec}^{-2},
\label{eqn:gal}
\end{equation}
and the probability of chance coincidence is therefore:
\begin{equation}
P(<\delta R)=1-{\rm e}^{-\pi (\delta R)^2\sigma(\le m_r)}\approx
0.02.
\label{eqn:prob}
\end{equation}
Given the low value of $P(<\delta R)$ and the absence of other
candidate hosts in the vicinity of the afterglow position, we consider
this galaxy to be the host of GRB\,111117A.  We note that with the XRT
position alone, the probability of chance coincidence for this galaxy
is much larger, $P(<\delta R)\approx 0.17$ (using $\delta R\approx
3\sigma_{\rm X,XRT}$; see \citealt{Bloom02}).

\subsection{Radio Observations}
\label{SubSec:RadioObs}

We observed the location of GRB\,111117A with the Karl G.~Jansky Very
Large Array \citep{rperley+11} on 2011 November 18.00 UT ($\delta t\approx 0.5$ d) at a
mean frequency of $5.8$ GHz with a total on-source integration time of
$75$ min.  We used 3C48 and J0042+2320 for bandpass/flux and gain
calibration, respectively, and followed standard procedures in the
Astronomical Image Processing System (AIPS, \citealt{Greisen03}) for data calibration and
analysis.  The effective bandwidth is about 1.5 GHz after excising
edge channels and data affected by radio frequency interference. 
We re-flagged and calibrated our data after the initial quick reduction
\citep{Fong11c} and
do not detect any significant emission in coincidence with the
\chandra\ position to a $3\sigma$ limit of $18$ $\mu$Jy.

\section{Results and Discussion}
\label{Sec:Disc}

\subsection{Host galaxy properties}
\label{Sec:Host}

\begin{figure}
\vskip -0.0 true cm
\centering
\includegraphics[scale=0.5]{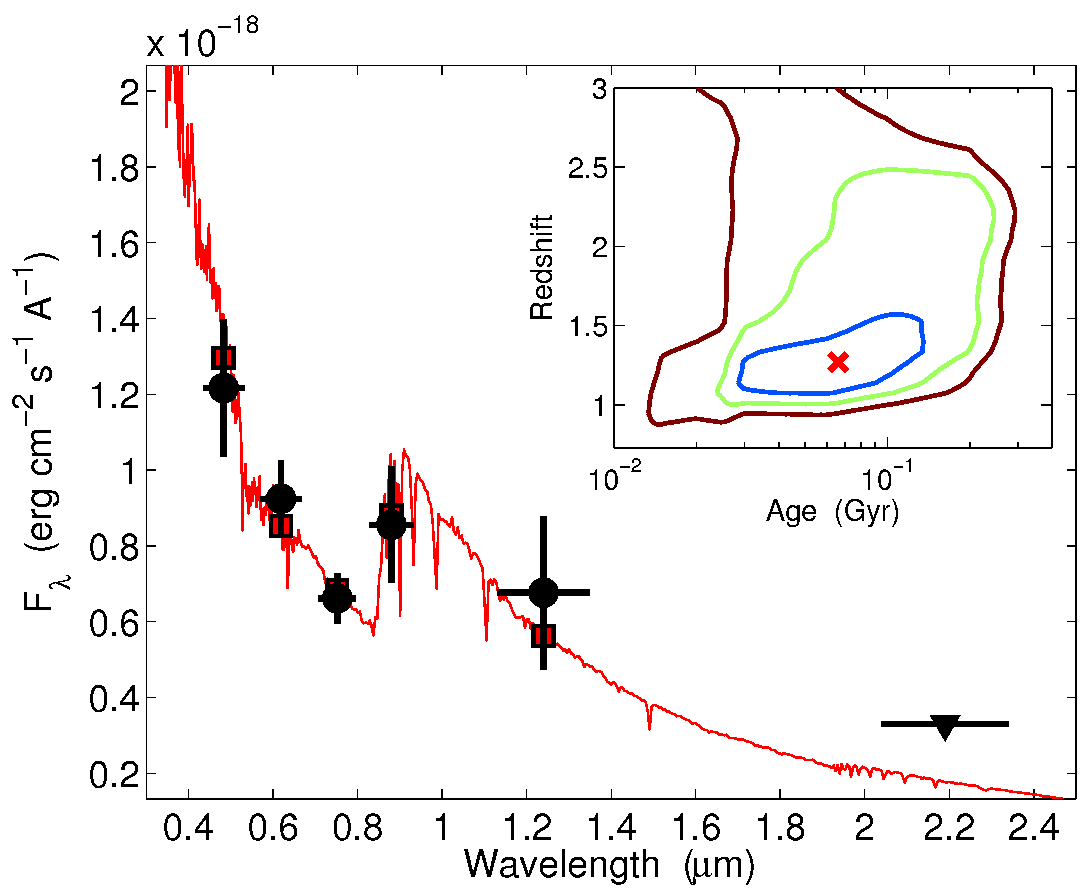}
\caption{Optical and near-IR spectral energy distribution of the host
galaxy of GRB\,111117A (black), along with the \citet{Maraston05}
evolutionary stellar population synthesis models.  The best-fit model
(red squares and line) has a photometric redshift of $z\approx 1.3$
and a stellar population age of $\tau\approx 70$ Myr (inset).}
\label{Fig:hostsed}
\end{figure}

To determine the photometric redshift and properties of the host
galaxy we use our $grizJK_s$ band photometry. We model
the host spectral energy distribution (SED) with the
\citet{Maraston05} evolutionary stellar population synthesis models,
using a Salpeter initial mass function, solar metallicity, and a red
horizontal branch morphology, with the redshift ($z$) and stellar
population age ($\tau$) as free parameters.  The resulting best-fit
model is shown in Figure~\ref{Fig:hostsed}, along with the confidence
regions for the redshift and age.  We find that $z=1.3^{+0.3}_{-0.2}$
and $\tau=70^{+65}_{-40}$ Myr ($\chi^2=1.2$ for $3$ d.o.f.); the
results remain unchanged if we use a model with a metallicity of 0.5
Z$_\odot$.  The inferred redshift is consistent with the
independent estimate by \cite{Sakamoto12} and is one of the highest for any short
GRB to date, either from spectroscopic or photometric measurements
\citep{Levan06b,Postigo06,Berger07,Graham09,Leibler10}, but is in the
range of expected redshifts for short GRBs with faint hosts
\citep{Berger07}. 

The inferred stellar population age is at the low end of the
distribution for short GRB hosts, for which $\langle\tau\rangle\approx
0.3$ Gyr \citep{Leibler10}.  The inferred host galaxy stellar mass is
$M_*\approx 4\times 10^9$ M$_\odot$, about a factor of 3 times lower
than the median for short GRB hosts, but this assumes a single stellar
population.  Contribution from an older stellar population could
increase the total mass up to a maximal value of $\approx 7\times
10^{10}$ M$_\odot$ if we assume the presence of a stellar population
with the age of the universe at $z=1.3$ (c.f., \citealt{Leibler10}).
We note that the stellar population age and specific star formation rate are 
similar to those of long GRB host galaxies, for which $\langle\tau\rangle\approx 
60$ Myr and $\langle{\rm SFR}/L_B\rangle\approx 10$ M$_\odot$ yr$^{-1}$ 
$L_B^{*-1}$ \citep{Berger09,Leibler10}.

From the observed $g$-band flux density, which samples the rest-frame
UV luminosity, we infer a star formation rate of ${\rm SFR}\approx 6$
M$_\odot$ yr$^{-1}$ \citep{Kennicutt98}.  This is higher than for most
previous short GRB host galaxies \citep{Berger09}.  The absolute
$B$-band magnitude is $M_B\approx -21.0$ mag, corresponding to
$L_B\approx 0.6$ L$^*$ in comparison to the DEEP2 luminosity function
at $z\approx 1.1$ \citep{Willmer06}; this value is typical for short
GRB hosts \citep{Berger09}.  Combining the star formation rate and
$B$-band luminosity, we infer a specific star formation rate of ${\rm
SFR}/L_B \approx 10$ M$_\odot$ yr$^{-1}$ $L_B^{*-1}$.  This is again
at the upper end of the distribution for short GRB host galaxies
\citep{Berger09}.

The host galaxy of \grb\ is overall similar to the host of short
GRB\,060801 ($z=1.130$) in terms of its star formation rate and
stellar mass \citep{Berger09}.  It provides additional support to the
conclusion that short GRB progenitors originate from diverse stellar
populations.  Under the assumption that the stellar population ages
can be used as a proxy for the progenitor delay time distribution
\citep{Leibler10}, events like \grb\ point to delay times as short as
a few tens of Myr.  In the context of NS-NS/NS-BH mergers, this is
suggestive of a subset of short-lived compact object binaries (e.g.,
\citealt{Belczynski01,Belczynski02b})

\subsection{Offset}
\label{Sec:Offset}

\begin{figure}
\vskip -0.0 true cm
\centering
\includegraphics[angle=0,scale=0.45]{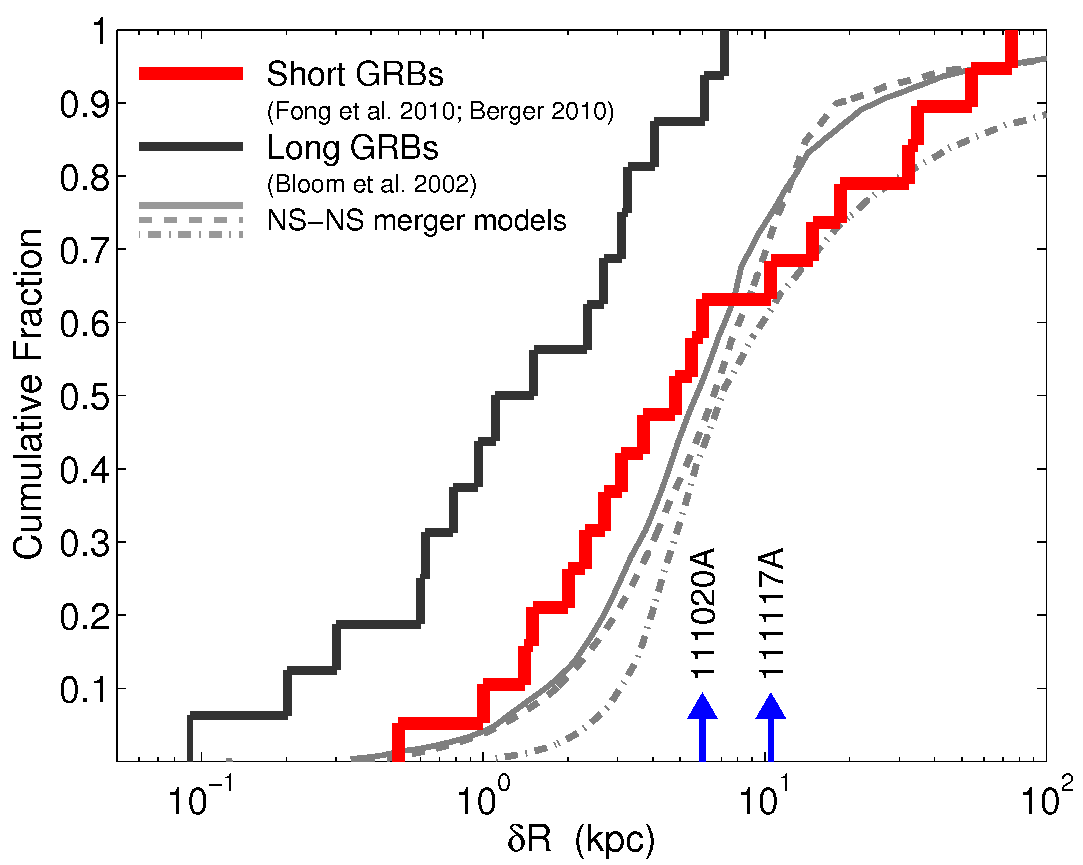}  
\caption{Cumulative distribution of projected physical offsets for
short GRBs with sub-arcsecond positions (red;
\citealt{Fong10,Berger10,Fong12}), including \grb, and for long GRBs
(black; \citealt{Bloom02}).  Also shown are population synthesis model
predictions for NS-NS binaries \citep{Bloom99,Fryer99,Belczynski06}.
Arrows mark the offsets of the two short GRBs localized by \chandra\
alone, GRB\,111117A from this work and GRB\,111020A from
\citet{Fong12}.  These offsets are somewhat larger than the median
short GRB offset of about 5 kpc.}
\label{Fig:offset}
\end{figure}

The \chandra-derived projected angular offset of $1.25\pm 0.20''$
corresponds to a projected physical offset of $\delta R=10.5\pm 1.7$
kpc at $z=1.3$.  This is comparable to the median offset of
about $5$ kpc for the sample of short GRBs studied by \citet{Fong10}
and \citet{Berger10}; see Figure~\ref{Fig:offset}).  Indeed, as a
subset, the two short bursts with precise localizations from \chandra\
alone (GRB\,111020A from \citealt{Fong12} and \grb\ presented here)
have similar offsets to those inferred from optical afterglows.  This
suggests that optical afterglows do not produce an obvious bias
against large offsets, as already demonstrated for the subset of short
GRBs that lack coincident host galaxies \citep{Berger10}.

Different short GRB progenitor models predict distinct offset
distributions.  NS-NS/NS-BH merger models predict a median offset of
about $5-10$ kpc \citep{Bloom99,Fryer99,Belczynski06} for host
galaxies with a mass comparable to the Milky Way as found for short
GRBs \citep{Berger09}.  On the other hand, magnetar models are not
expected to produce a substantial fraction of short GRBs at offsets of
$\gtrsim 10$ kpc \citep{Levan06,Metzger08}.  Figure~\ref{Fig:offset}
shows that NS-NS binary models are in reasonable agreement with the
observed distribution of physical offsets, from both optical and
\chandra\ positions.

\subsection{X-ray afterglow properties }
\label{Sec:XrayAft}

\begin{figure*}
\vskip -0.0 true cm
\centering
\includegraphics[scale=0.7]{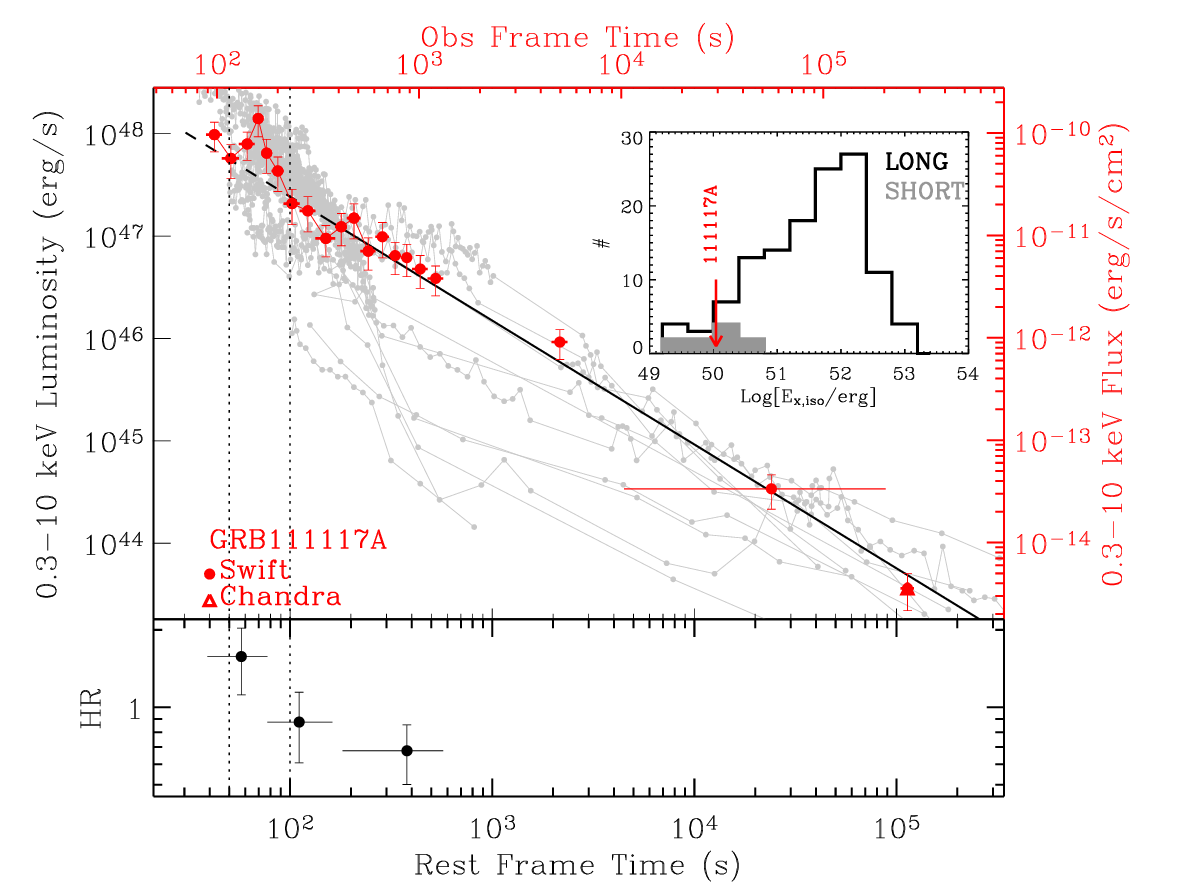}
\caption{{\it Upper panel:} Unabsorbed $0.3-10$ keV flux and
luminosity light curve of GRB\,111117A (red dots: \swift/XRT; red
triangle: \chandra) compared to 11 short GRBs detected by Swift for
which a redshift measurement is available (grey lines).  The best fit
power-law model has $\alpha_x=-1.21\pm 0.05$ (solid black line).  An
apparent flare is detected at $\approx 80-200$ s (black dotted lines).
The inset shows the distribution of the isotropic energy emitted
during the X-ray afterglow for long GRBs (black line) and short GRBs
(grey filled histogram) as computed by \citet{Margutti12}.  For
GRB\,111117A we measure $E_{\rm x,iso}\approx 1.1\times 10^{50}$ erg
s$^{-1}$.  {\it Lower panel:} Time evolution of the hardness ratio
measured between the $1.5-10$ keV and the $0.3-1.5$ keV energy bands.}
\label{Fig:XRTlc}
\end{figure*}

\begin{figure*}
\vskip -0.0 true cm
\centering
\includegraphics[scale=0.4]{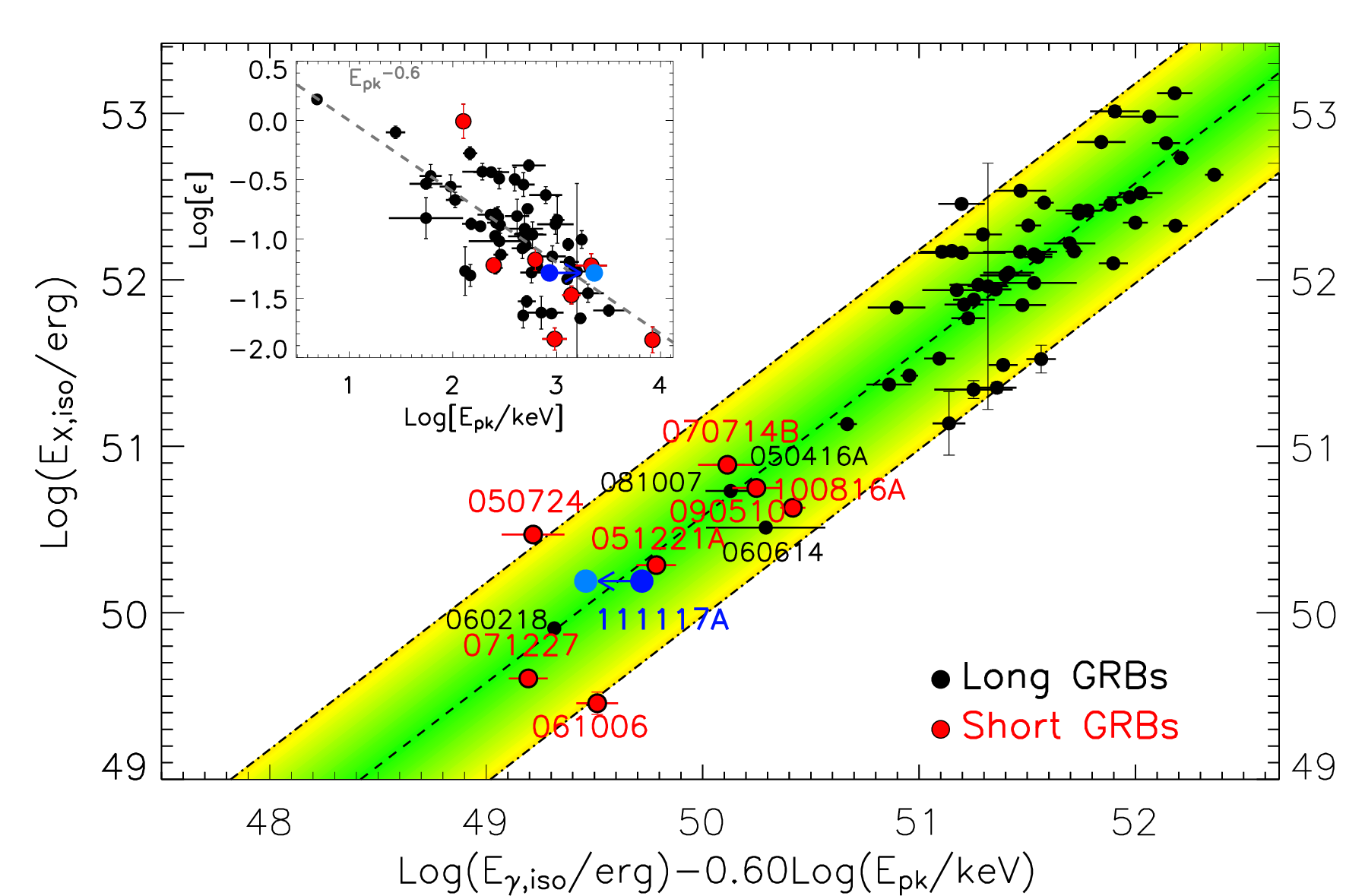}
\caption{Three-parameter correlation involving the isotropic energy
emitted in the X-ray afterglow ($E_{\rm{x,iso}}$; rest-frame $0.3-30$
keV), the isotropic $\gamma$-ray energy ($E_{\rm\gamma,iso}$;
rest-frame $1-10^4$ keV), and the rest-frame spectral peak energy
during the prompt phase ($E_{\rm{pk}}$).  The blue circles mark
GRB\,111117A using the range of $E_{\rm{pk}}\sim 850-2300$ keV
(\S\ref{SubSec:BAT}).  The dot-dashed lines mark the 90\% confidence
area around the best-fit relation: $E_{\rm x,iso}\propto
E_{\rm\gamma,iso}^{1.00\pm 0.06}\,E_{\rm pk}^{-0.60\pm 0.10}$.  The
inset shows the evolution of $\epsilon\equiv E_{\rm x,iso}/E_{\rm
\gamma,iso}$ as a function of $E_{\rm{pk}}$.  Adapted from
\citet{Margutti12}.}
\label{Fig:3parcorr}
\end{figure*}

\begin{figure*}
\vskip -0.0 true cm
\centering
\includegraphics[scale=0.5]{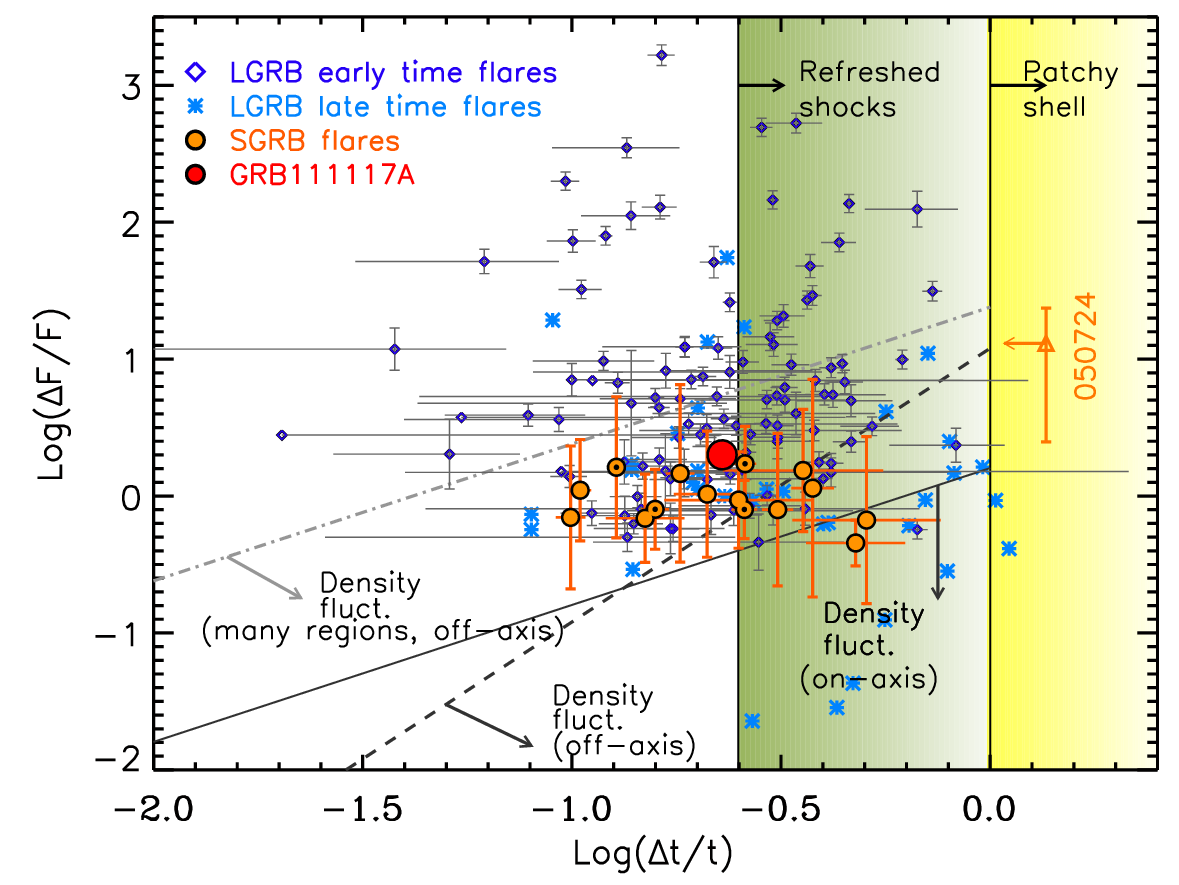}
\caption{Relative variability flux ($\Delta F/F$) versus relative
variability time scale ($\Delta t/t$) for a sample of short GRB X-ray
flare candidates (orange filled circles; \citealt{Margutti11}), as
well as early (blue open diamonds; \citealt{Chincarini10}) and
late-time (light-blue stars; \citealt{Bernardini11}) flares in long
GRBs.  A small black dot marks short GRBs with extended emission.  The
apparent flare detected in \grb\ is marked with a red filled circle.
The late-time re-brightening detected in GRB\,050724 is also shown for
completeness with an orange open triangle.  Solid, dashed and
dot-dashed lines mark the kinematically allowed regions according to
\citet{Ioka05} (their equations (7) and (A2)).}
\label{Fig:Ioka}
\end{figure*}

At $z=1.3$ the X-ray afterglow of GRB\,111117A lies at the upper end
of the short GRB luminosity distribution, with a typical power-law
decay (Figure~\ref{Fig:XRTlc}).  The total energy released in the
$0.3-10$ keV energy band during the X-ray afterglow ($80$ s to $3$ d)
is $E_{\rm x,iso}=(1.1\pm 0.1)\times 10^{50}$ erg, typical
for short GRBs (Figure~\ref{Fig:XRTlc} inset).  This confirms previous
findings that the X-ray afterglows of short GRBs are on average $\sim
100$ times less energetic than those of long GRBs \citep{Margutti12}.
The corresponding energy radiated in the $0.3-30$ keV rest frame
band\footnotemark\footnotetext{This value is obtained by extrapolating
the spectral behavior in the $0.3-10$ keV observer frame range
($0.7-23$ keV rest frame) to the $0.3-30$ keV rest frame range.} is
$E_{\rm x,iso}\approx 1.5\times 10^{50}$ erg.  In comparison
to the isotropic $\gamma$-ray energy this indicates $E_{\rm
x,iso}\approx 0.03 E_{\rm\gamma,iso}$, which is typical of short
GRBs\footnotemark\footnotetext{We assumed a high energy photon index
$\beta=-2.4$ (see e.g. \citealt{Kaneko06}) to extrapolate the 23-2300
keV (rest-frame) spectrum to the $1-10^4$ keV rest-frame range.}

We combine this result with the inferred rest frame value of $E_{\rm
pk}\sim 850-2300$ keV to show that \grb\ is consistent with the
recently-reported $E_{\rm x,iso}-E_{\rm\gamma,iso}-E_{\rm pk}$
correlation for long and short GRBs, and resides in the region
populated by short GRBs and X-ray flashes (Figure~\ref{Fig:3parcorr};
\citealt{Margutti12}).  This provides additional support
to the conclusions that: (i) this correlation can be used to divide
``standard'' long GRBs from short GRBs, peculiar GRBs, and X-ray
flashes; and (ii) the physical origin of the correlation is related to
a common feature of the different classes, possibly the properties of
the relativistic outflow (in particular the bulk Lorentz factor; see
\citealt{Bernardini12} for details; see \citealt{Fan12} and
\citealt{Dado12} for alternative explanations).

At $\delta t\approx 80-250$ s we find evidence for an apparent flare
superimposed on the smooth X-ray afterglow decay, with a significance
of $\approx 3\sigma$ (Figure~\ref{Fig:XRTlc}).  With a rest frame
duration of $\Delta t\approx 16$ s and peak time of $t_{\rm pk}\approx
70$ s, the flare is consistent with the $\Delta t$ versus $t_{\rm pk}$
correlation established by long GRB flares \citep{Chincarini10} and
shared by short GRBs flares \citep{Margutti11}.  The flux contrast of
the flare, $\Delta F/F\approx 2$, is also typical of flares in short
GRBs (Figure~\ref{Fig:Ioka}).  Finally, the flare peak luminosity and
integrated energy are $L_{\rm pk}^{\rm flare}\approx 10^{48}$ erg
s$^{-1}$ and $E_{\rm X}^{\rm flare}\approx 1.6\times 10^{49}$ erg,
again typical of short GRB flares \citep{Margutti11}.  We note that
the value of $\Delta t/t_{\rm pk}\approx 0.2$ does not support an
external shock origin, for which we expect $\Delta t/t_{\rm pk}\gtrsim
1$ (e.g., \citealt{Zhang06}; but see \citealt{Dermer08}).  However,
the flux contrast of $\Delta F/F\approx 2$ is also at odds with the
expectation for central engine variability, with $\Delta F/F\approx
100$ \citep{Lazzati11}.

\subsection{Multi-wavelength Afterglow Modeling}
\label{Sec:Aft}

\begin{figure*}
\vskip -0.0 true cm
\centering
\includegraphics[angle=0,width=2.2in]{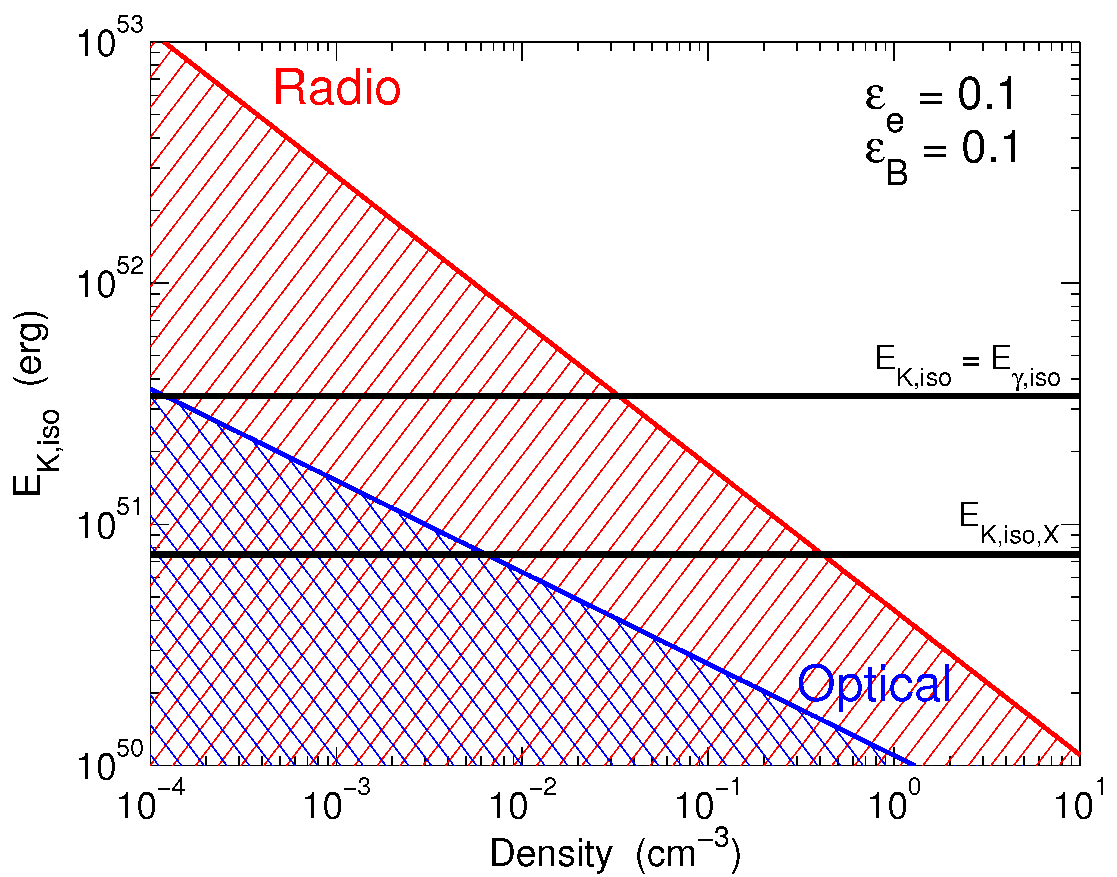}
\includegraphics[angle=0,width=2.2in]{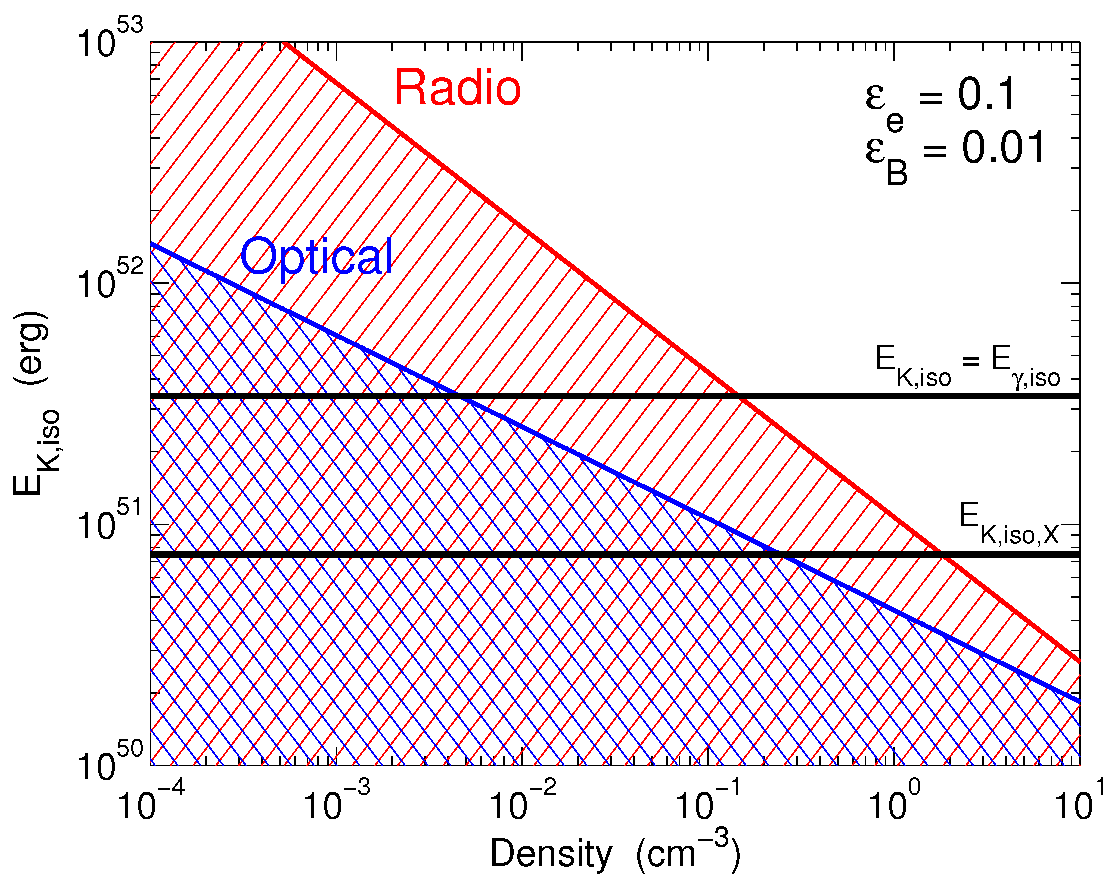}
\includegraphics[angle=0,width=2.2in]{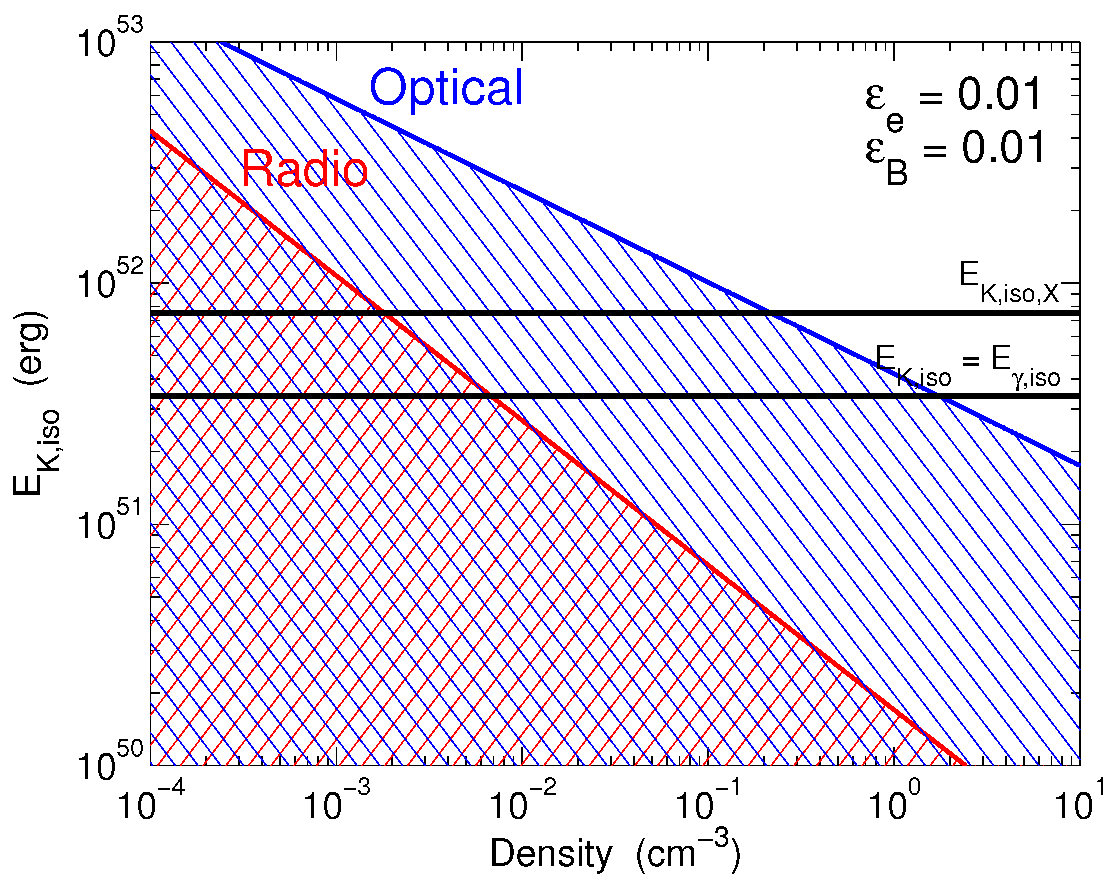}\\
\includegraphics[angle=0,width=2.2in]{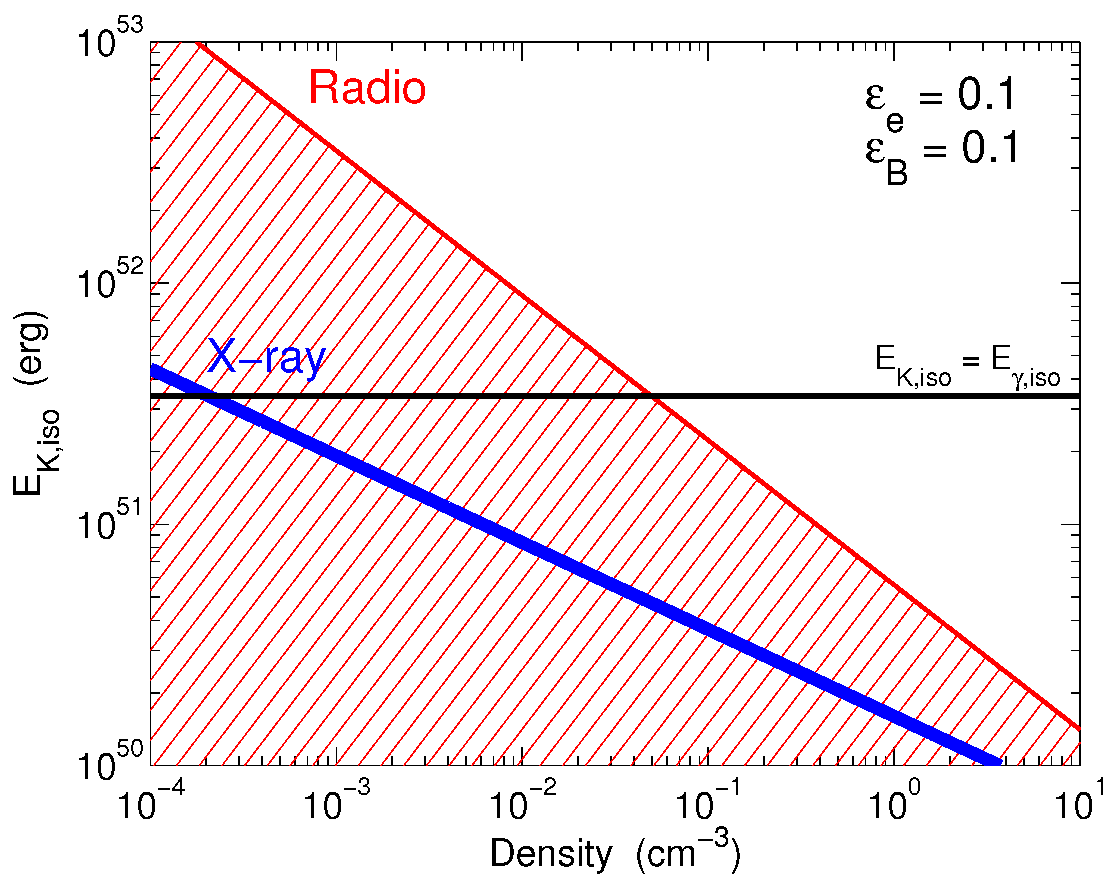}
\includegraphics[angle=0,width=2.2in]{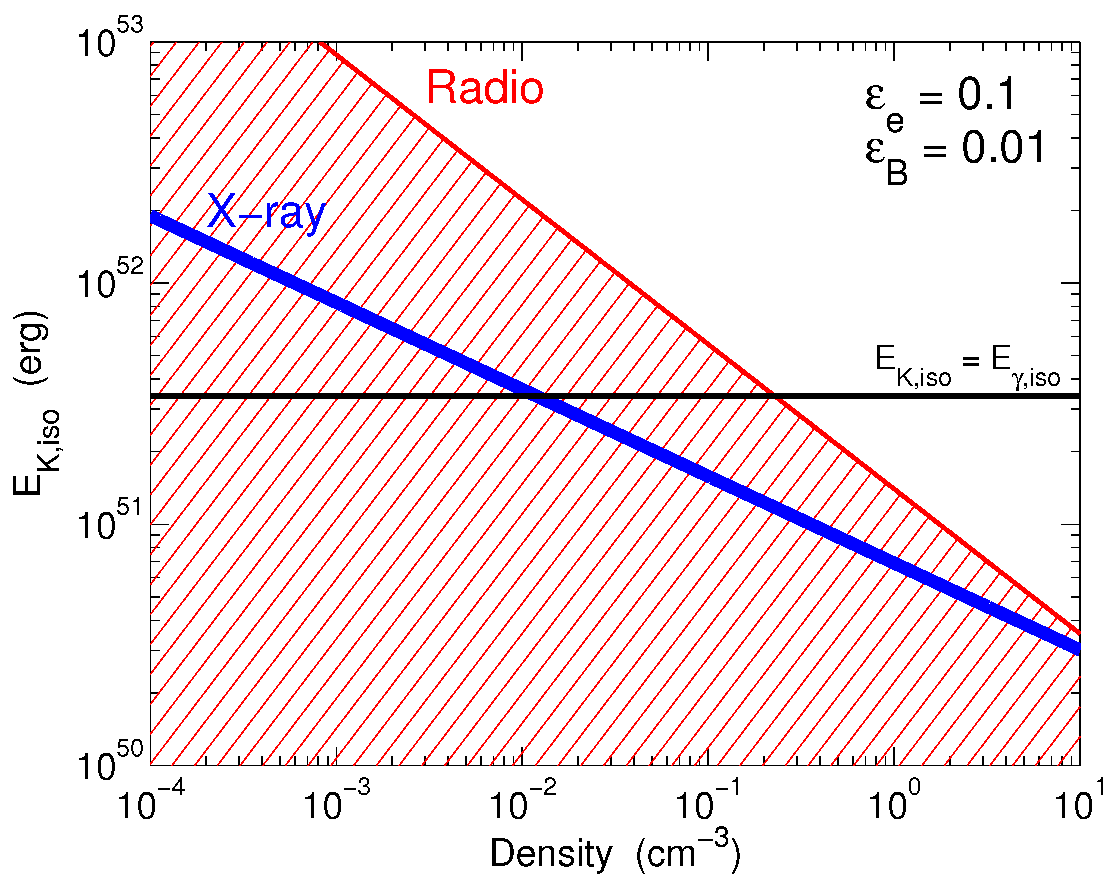}
\includegraphics[angle=0,width=2.2in]{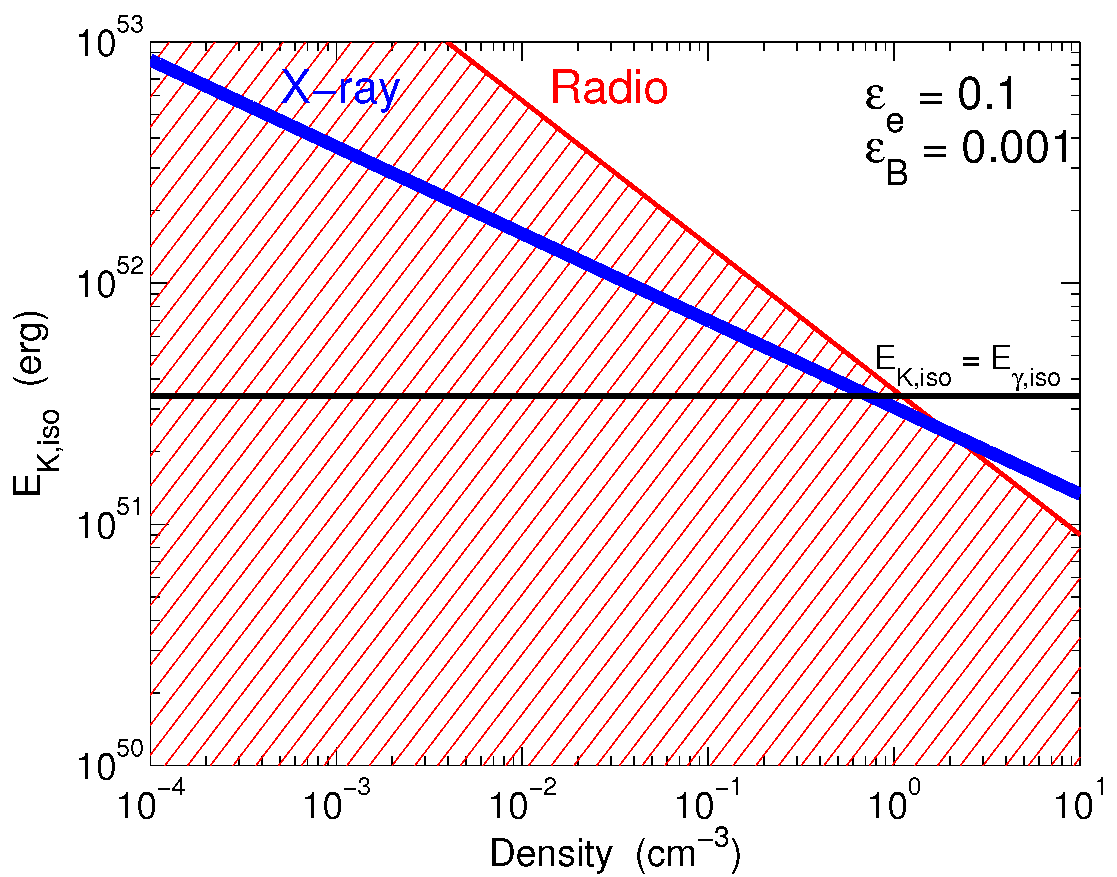}     
\caption{The phase-space of isotropic-equivalent blastwave kinetic
energy and circumburst density delineated by the X-ray detection
(thick line) and the optical and/or radio upper limits (hatched
regions).  The top row is for the case $\nu_c<\nu_X$, with the X-ray
detection providing a measure of the $E_{\rm K,iso}$ (horizontal
line).  The bottom row is for the case $\nu_c>\nu_X$ for which the
optical limit is redundant with respect to the X-ray detection.  Here
the case of $E_{\rm K,iso}=E_{\rm\gamma,iso}$ is marked by the
horizontal black line.  The panels are for different combinations of
$\epsilon_e$ and $\epsilon_B$.}
\label{fig:afterglow}
\end{figure*}

The detected X-ray afterglow, along with the upper limits in the
optical and radio allow us to extract some of the basic properties of
GRB\,111117A.  We adopt the afterglow synchrotron model formulation of
\citet{Granot02}, which provides a mapping from the observed fluxes
and break frequencies to the isotropic-equivalent kinetic energy
($E_{\rm K,iso}$), circumburst density ($n_0$), the fractions of
post-shock energy in radiating electrons ($\epsilon_e$) and magnetic
fields ($\epsilon_B$), and the electron power-law distribution index
($p$, with $N(\gamma)\propto\gamma^{-p}$).  We consider the case of a
constant density medium relevant for short GRBs.  In the analysis
below we adopt the best-fit redshift of $z=1.3$.

The X-ray temporal and spectral indices are $\alpha_X=-1.21\pm 0.05$
and $\beta_X=-1.0\pm 0.2$ (\S\ref{SubSec:XObs}).  For the case of the
synchrotron cooling break located redward of the X-ray band
($\nu_c<\nu_X$), the resulting values of $p$ are $2.28\pm 0.07$ and
$2.0\pm 0.4$ from $\alpha_X$ and $\beta_X$, respectively; for the
opposite scenario ($\nu_c>\nu_X$) the resulting values of $p$ are
$2.61\pm 0.07$ and $3.0\pm 0.4$.  In both cases the values of $p$
inferred from $\alpha_X$ and $\beta_X$ are consistent, indicating that
the X-ray data alone cannot distinguish the location of $\nu_c$.

The unabsorbed X-ray flux density at the time of the optical
non-detection ($\delta t\approx 0.55$ d) is $F_{\rm \nu,X}\approx 4.4$
nJy, compared to $F_{\rm \nu,opt}\lesssim 0.23$ $\mu$Jy.  This leads
to an observed spectral index of $\beta_{\rm OX}\gtrsim -0.63$,
consistent with the value of $p\approx 2.28$ for the case of
$\nu_c<\nu_X$, if $\nu_c\approx\nu_X$ (i.e., if the relevant spectral
slope between the X-ray and optical bands is $\beta_{OX}=-(p-1)/2$).
On the other hand, if $\nu_c>\nu_X$ we expect a spectral index of
$\beta_{OX}=-(p-1)/2\approx -0.80$, which is much steeper than the
observed value.  With this spectral index we would expect the optical
flux to be $\approx 0.65$ $\mu$Jy, or about 1.1 mag brighter than the
observed limit.  Thus, the X-ray/optical comparison either requires
rest-frame extinction of $A_V^{\rm host}\gtrsim 0.5$ mag or
$\nu_c\approx\nu_X$.  We note that for the Galactic relations between
$N_H$ and $A_V$ \citep{Predehl95} the optical extinction would imply
$N_H^{\rm host}\gtrsim 10^{21}$ cm$^{-2}$, consistent with the
marginal detection in the XRT spectrum of $(6.7\pm 3.0)\times 10^{21}$
cm$^{-2}$.

Under the assumption that $\nu_c<\nu_X$ we can use the \chandra\ X-ray
flux density, $F_{\rm\nu,X}\approx 0.42$ nJy, to infer the value of
the isotropic-equivalent kinetic energy:
\begin{equation}
E_{\rm K,iso}\approx 7.5\times 10^{50}\,\,\,{\rm erg},
\label{eqn:ekiso}
\end{equation} 
where we have assumed $\epsilon_e=\epsilon_B=0.1$; we note that the
dependence on $\epsilon_B$ is weak, $E_{\rm K,iso}\propto
\epsilon_B^{-0.07}$, while $E_{\rm K,iso}$ is inversely proportional
to $\epsilon_e$.  The fiducial value of $E_{\rm K,iso}$ is lower than
the isotropic-equivalent $\gamma$-ray energy, $E_{\rm\gamma,iso}
\approx 3.0\times 10^{51}$ erg.

We next use the upper bounds on the radio and optical flux densities
to place constraints on the circumburst density and energy.  For the
radio upper limit we use the synchrotron flux density relevant for
$\nu_a<\nu_{\rm rad}<\nu_m$:
\begin{equation} 
F_{\rm\nu,rad}\approx 36\,{\rm\mu Jy}\,n_0^{1/2}\,E_{\rm
K,iso,51}^{5/6}\lesssim 18\,{\rm\mu Jy},
\end{equation}
while for the optical upper limit we use the synchrotron flux density
relevant for $\nu_m<\nu_{\rm opt}<\nu_c$:
\begin{equation}
F_{\rm\nu,opt}\approx 4.4\,{\rm\mu Jy}\,n_0^{1/2}\,E_{\rm
K,iso,51}^{1.32}\lesssim 0.23\,{\rm\mu Jy}.
\end{equation}
The resulting allowed phase-space of $E_{\rm K,iso}$ and $n_0$ is
shown in Figure~\ref{fig:afterglow}.  Using the value of $E_{\rm
K,iso}$ inferred from the X-ray data (Equation~\ref{eqn:ekiso}), and
the corresponding limits on $n_0$ from the optical data ($n_0\lesssim
0.006$ cm$^{-3}$) and radio data ($n_0\lesssim 0.4$ cm$^{-3}$), we
find that the cooling frequency is located at $\nu_c\gtrsim
(0.15-8)\times 10^{17}$ Hz (i.e., $\gtrsim 0.06-3$ keV) at $\delta
t=1000$ s.  Using instead a value of $\epsilon_B=0.01$ the limits on
the density are $\lesssim 0.25$ cm$^{-3}$ (optical) and $\lesssim 2$
cm$^{-3}$ (radio), and the cooling frequency is therefore
$\nu_c\gtrsim (0.9-7)\times 10^{17}$ Hz (i.e., $\gtrsim 0.4-3$ keV).
Thus, the inferred location of $\nu_c$ is in the X-ray band, in
agreement with our conclusion from the comparison of X-ray and optical
flux densities.

In the alternative scenario of $\nu_c>\nu_X$ both the optical and
X-ray bands probe the same portion of the synchrotron spectrum,
$\nu_m<\nu_{\rm opt,X}<\nu_c$, but this time with a value of $p=2.61$.
For the X-ray band, this gives the relation (for $\epsilon_e=
\epsilon_B=0.1$):
\begin{equation}
F_{\rm\nu,X}\approx 4.4\,{\rm nJy}\,n_0^{1/2}\,E_{\rm K,iso,51}^{1.4}
\approx 0.42\,{\rm nJy},
\end{equation}
while for the radio band ($\nu_a<\nu_{\rm rad}<\nu_m$) we find:
\begin{equation} 
F_{\rm\nu,rad}\approx 12\,{\rm\mu Jy}\,n_0^{1/2}\,E_{\rm
K,iso,51}^{5/6}\lesssim 18\,{\rm\mu Jy}.
\end{equation}

The resulting allowed regions of $E_{\rm K,iso}-n_0$ phase-space are
shown in Figure~\ref{fig:afterglow}.  Assuming that $E_{\rm K,iso}=
E_{\rm \gamma,iso}=3.0\times 10^{51}$ erg, the X-ray flux density
corresponds to $n_0\approx 3\times 10^{-4}$ cm$^{-3}$; for
$\epsilon_B=0.01$ the density is instead\footnotemark\footnotetext{We
can rule out $\epsilon_e=\epsilon_B=0.01$ or $\epsilon_e=0.1$ and
$\epsilon_B\lesssim 0.001$ since in these cases the upper limit on the
density from the radio data is lower than the density inferred from
the X-ray detections.} $\approx 0.02$ cm$^{-3}$, while for
$\epsilon_B=0.001$ the density is $\approx 1.2$ cm$^{-3}$.  With these
values we indeed find that $\nu_c\gtrsim 4\times 10^{18}$ Hz ($\gtrsim
16$ keV) at $\delta t=1000$ s, consistent with the assumption that
$\nu_c>\nu_X$.

To conclude, with the assumption that $\nu_c<\nu_X$ we find that
$E_{\rm K,iso}\approx 7.5\times 10^{50}$ erg, and $n_0\lesssim 0.01$
cm$^{-3}$ ($\epsilon_B=0.1$) or $\lesssim 0.2$ cm$^{-3}$
($\epsilon_B=0.01$).  The resulting location of the cooling break
indicates that $\nu_c\sim\nu_X$, marginally consistent with the
inherent assumption.  On the other hand, if $\nu_c>\nu_X$, then the
assumption of $E_{\rm K,iso}\approx E_{\rm\gamma,iso}$ indicates that
$n_0\approx 3\times 10^{-4}-1$ cm$^{-3}$ (for $\epsilon_B=0.001-0.1$).
However, this also requires host galaxy extinction with $A_V^{\rm
host}\gtrsim 0.5$ mag.  In both cases the inferred density is
consistent with the NS-NS/NS-BH merger scenario, for which the
expected densities are $n_0\sim 10^{-6}-1$ cm$^{-3}$
\citep{Perna02,Belczynski06}.

Finally, the non-detection of a break in the X-ray light curve to
$\delta t\approx 3$ d allows us to place a lower limit on the opening
angle of the outflow from GRB\,111117A.  Using the formulation of
\citep{Frail01}\footnotemark\footnotetext{In this calculation the
observer is assumed to be on-axis.} we find that for $E_{\rm K,iso}=
E_{\rm \gamma,iso}$ and $n_0\approx 3\times 10^{-4}-1$ cm$^{-3}$, the
resulting lower limit is $\theta_j\gtrsim 3-10^\circ$.  This range
indicates a beaming correction as low as $\approx 70$ and as high as
about $600$.

\section{Summary and Conclusions}
\label{Sec:Conc}

We presented multi-wavelength observations of the afterglow of short
GRB\,111117A, along with optical and near-IR follow-up observations of
its host galaxy.  These observations provide critical insight into the
nature of \grb:

\begin{itemize} 

\item Using a \chandra\ observation we accurately pinpoint the
location of the afterglow to the outskirts of a galaxy at a
photometric redshift of $z\approx 1.3$, one of the highest for any
short GRB to date.  The projected physical offset is about $10.5$ kpc,
reminiscent of previous short GRBs \citep{Fong10,Berger10}.  Along
with the previous burst detected by \chandra\ alone (GRB\,111020A;
\citealt{Fong12}) we find that short GRBs localized by optical and
X-ray afterglows appear to have similar offsets.

\item The host galaxy of GRB\,111117A exhibits vigorous star formation
activity and a young stellar population age that are at the upper
bound of the distribution for short GRB hosts
\citep{Berger09,Leibler10}.

\item The X-ray afterglow properties are typical of short GRBs with
long-lasting X-ray emission.  In particular, with $E_{\rm
x,iso}\approx 1.5 \times 10^{50}$ erg, $E_{\rm\gamma,iso}\approx
3\times 10^{51}$ erg, and $E_{\rm pk}\approx 850-2300$ keV,
GRB\,111117A is consistent with the three-parameter universal GRB scaling
recently reported by \citet{Margutti12}.  The X-ray to $\gamma$-ray
energy ratio for \grb\ is $\epsilon\approx 0.03$, as typically found
for short GRBs.

\item We find evidence (statistical significance of $\sim3\sigma$) for an early flare superimposed on the X-ray
afterglow decay with properties that are reminiscent of X-ray flare
candidates detected in other short GRBs \citep{Margutti11}.  The
origin of X-ray flares appears to be independent of the large scale
environment since they are detected from short GRBs in both early- and
late-type galaxies.

\item Using the X-ray light curve, and deep upper limits in the
optical and radio bands we find that if $\nu_c>\nu_X$ then
$n_0^{1/2}\,E_{\rm K,iso,51}^{1.4}\approx 0.1-6$ (for $\epsilon_e=0.1$
and $\epsilon_B=0.001-0.1$).  For the specific case of $E_{\rm K,iso}=
E_{\rm \gamma,iso}\approx 3.0\times 10^{51}$ erg, this leads to a
density of $n_0\approx 3\times 10^{-4}-1$ cm$^{-3}$; larger densities
are ruled out independently by the radio limit, which leads to
$n_0^{1/2}\,E_{\rm K,iso,51}^{5/6}\lesssim 0.7-3$ (for
$\epsilon_e=0.1$ and $\epsilon_B=0.001-0.1$).  However, this scenario
requires substantial rest-frame extinction of $A_V^{\rm host}\gtrsim
0.5$ mag to explain the optical non-detection.  In the alternative
scenario of $\nu_c<\nu_X$ we find that $E_{\rm K,iso}\approx 7.5\times
10^{50}\,(\epsilon_e/0.1)^{-1}$ erg and $n_0\lesssim 0.1$ cm$^{-3}$.

\item The lack of a clear break in the X-ray light curve at $\lesssim
3$ d, points to an opening angle of $\theta_j\gtrsim 3-10^\circ$, with
the exact lower limit depending on the circumburst density.

\end{itemize}

The results of this work highlight the importance of \chandra\ for the
determination of short GRB sub-arcsecond positions, especially in the
absence of optical detections.  This is critical for locating short
GRBs within their host environments, particularly in comparison to
\swift/XRT position, which are generally much larger than the host
galaxy sizes.

\acknowledgments R.M.~thanks Rodolfo Barniol Duran and Cristiano
Guidorzi for useful discussions.  We thank Francesco di Mille for
carrying out the Magellan FourStar observations, and Andy Monson for
assistance with the data reduction.  The Berger GRB group at Harvard
is supported by the National Science Foundation under Grant
AST-1107973.  Partial support was also provided by NASA/Swift AO6
grant NNX10AI24G. Support for this work was provided by the David and 
Lucile Packard Foundation Fellowship for Science and Engineering  awarded to A.M.S
S.B.C.~acknowledges generous financial assistance 
from Gary \& Cynthia Bengier, the Richard \& Rhoda Goldman Fund, 
NASA/{\it Swift} grants NNX10AI21G and NNX12AD73G, the TABASGO 
Foundation, and US National Science Foundation (NSF) grant AST-0908886.
A.R. and S.K. acknowledge support by grant DFG Kl 766/16-1. S.S.
 acknowledges support by a Landesstipendium. Part of the funding for
 GROND (both hardware and personnel) was generously granted by the
 Leibniz-Prize to G. Hasinger (DFG grant HA 1850/28-1).
Observations were obtained with the VLA (program
10C-145) operated by the National Radio Astronomy Observatory.
The National Radio Astronomy Observatory is a facility of the National Science 
Foundation operated under cooperative agreement by Associated Universities, Inc.
The paper includes data
gathered with the 6.5 meter Magellan Telescopes located at Las
Campanas Observatory, Chile.  This work is based in part on
observations obtained at the Gemini Observatory, which is operated by
the Association of Universities for Research in Astronomy, Inc., under
a cooperative agreement with the NSF on behalf of the Gemini
partnership: the National Science Foundation (United States), the
Science and Technology Facilities Council (United Kingdom), the
National Research Council (Canada), CONICYT (Chile), the Australian
Research Council (Australia), Ministério da Ciência, Tecnologia e
Inovação (Brazil) and Ministerio de Ciencia, Tecnología e Innovación
Productiva (Argentina).


\begin{thebibliography}{}

\bibitem[\protect\citeauthoryear{{Abazajian} et~al.}{{Abazajian}
  et~al.}{2009}]{Abazajian09}
{Abazajian}, K.~N., et~al. 2009, \apjs, 182, 543

\bibitem[\protect\citeauthoryear{{Andersen} et~al.}{{Andersen}
  et~al.}{2011}]{Andersen11}
{Andersen}, M.~I., {de Ugarte Postigo}, A., {Leloudas}, G.,  \& {Fynbo},
  J.~P.~U. 2011, GRB Coordinates Network, 12563, 1

\bibitem[\protect\citeauthoryear{{Barthelmy} et~al.}{{Barthelmy}
  et~al.}{2005}]{Barthelmy05}
{Barthelmy}, S.~D., et~al. 2005, \ssr, 120, 143

\bibitem[\protect\citeauthoryear{{Beckwith} et~al.}{{Beckwith}
  et~al.}{2006}]{Beckwith06}
{Beckwith}, S.~V.~W., et~al. 2006, \aj, 132, 1729

\bibitem[\protect\citeauthoryear{{Belczynski} \& {Kalogera}}{{Belczynski} \&
  {Kalogera}}{2001}]{Belczynski01}
{Belczynski}, K.,  \& {Kalogera}, V. 2001, \apjl, 550, L183

\bibitem[\protect\citeauthoryear{{Belczynski}, {Kalogera}, \&
  {Bulik}}{{Belczynski} et~al.}{2002}]{Belczynski02b}
{Belczynski}, K., {Kalogera}, V.,  \& {Bulik}, T. 2002, \apj, 572, 407

\bibitem[\protect\citeauthoryear{{Belczynski} et~al.}{{Belczynski}
  et~al.}{2006}]{Belczynski06}
{Belczynski}, K., {Perna}, R., {Bulik}, T., {Kalogera}, V., {Ivanova}, N.,  \&
  {Lamb}, D.~Q. 2006, \apj, 648, 1110

\bibitem[\protect\citeauthoryear{{Berger}}{{Berger}}{2009}]{Berger09}
{Berger}, E. 2009, \apj, 690, 231

\bibitem[\protect\citeauthoryear{{Berger}}{{Berger}}{2010}]{Berger10}
{Berger}, E. 2010, \apj, 722, 1946

\bibitem[\protect\citeauthoryear{{Berger}}{{Berger}}{2011}]{Berger11b}
{Berger}, E. 2011, New Astronomy Reviews, 55, 1

\bibitem[\protect\citeauthoryear{{Berger}, {Fong}, \& {Sakamoto}}{{Berger}
  et~al.}{2011}]{Berger11}
{Berger}, E., {Fong}, W.,  \& {Sakamoto}, T. 2011, GRB Coordinates Network,
  12588, 1

\bibitem[\protect\citeauthoryear{{Berger} et~al.}{{Berger}
  et~al.}{2007}]{Berger07}
{Berger}, E., et~al. 2007, \apj, 664, 1000

\bibitem[\protect\citeauthoryear{{Berger} et~al.}{{Berger}
  et~al.}{2005}]{Berger05}
{Berger}, E., et~al. 2005, \nat, 438, 988

\bibitem[\protect\citeauthoryear{{Bernardini} et~al.}{{Bernardini}
  et~al.}{2011}]{Bernardini11}
{Bernardini}, M.~G., {Margutti}, R., {Chincarini}, G., {Guidorzi}, C.,  \&
  {Mao}, J. 2011, \aap, 526, A27

\bibitem[\protect\citeauthoryear{{Bernardini} et~al.}{{Bernardini}
  et~al.}{2012}]{Bernardini12}
{Bernardini}, M.~G., {Margutti}, R., {Zaninoni}, E.,  \& {Chincarini}, G. 2012,
  arXiv:1203.1060

\bibitem[\protect\citeauthoryear{{Bloom}, {Kulkarni}, \& {Djorgovski}}{{Bloom}
  et~al.}{2002}]{Bloom02}
{Bloom}, J.~S., {Kulkarni}, S.~R.,  \& {Djorgovski}, S.~G. 2002, \aj, 123, 1111

\bibitem[\protect\citeauthoryear{{Bloom}, {Sigurdsson}, \& {Pols}}{{Bloom}
  et~al.}{1999}]{Bloom99}
{Bloom}, J.~S., {Sigurdsson}, S.,  \& {Pols}, O.~R. 1999, \mnras, 305, 763

\bibitem[\protect\citeauthoryear{{Burrows} et~al.}{{Burrows}
  et~al.}{2006}]{Burrows06}
{Burrows}, D.~N., et~al. 2006, \apj, 653, 468

\bibitem[\protect\citeauthoryear{{Burrows} et~al.}{{Burrows}
  et~al.}{2005}]{Burrows05}
{Burrows}, D.~N., et~al. 2005, \ssr, 120, 165

\bibitem[\protect\citeauthoryear{{Cenko} \& {Cucchiara}}{{Cenko} \&
  {Cucchiara}}{2011}]{Cenko11}
{Cenko}, S.~B.,  \& {Cucchiara}, A. 2011, GRB Coordinates Network, 12577, 1

\bibitem[\protect\citeauthoryear{{Chincarini} et~al.}{{Chincarini}
  et~al.}{2010}]{Chincarini10}
{Chincarini}, G., et~al. 2010, \mnras, 406, 2113

\bibitem[\protect\citeauthoryear{{Dado} \& {Dar}}{{Dado} \&
  {Dar}}{2012}]{Dado12}
{Dado}, S.,  \& {Dar}, A. 2012, arXiv:1203.5886

\bibitem[\protect\citeauthoryear{{de Ugarte Postigo} et~al.}{{de Ugarte
  Postigo} et~al.}{2006}]{Postigo06}
{de Ugarte Postigo}, A., et~al. 2006, \apjl, 648, L83

\bibitem[\protect\citeauthoryear{{Dermer}}{{Dermer}}{2008}]{Dermer08}
{Dermer}, C.~D. 2008, \apj, 684, 430

\bibitem[\protect\citeauthoryear{{Fan} et~al.}{{Fan} et~al.}{2012}]{Fan12}
{Fan}, Y.-Z., {Wei}, D.-M., {Zhang}, F.-W.,  \& {Zhang}, B.-B. 2012, arXiv:1204.4881

\bibitem[\protect\citeauthoryear{{Foley} \& {Jenke}}{{Foley} \&
  {Jenke}}{2011}]{Foley11}
{Foley}, S.,  \& {Jenke}, P. 2011, GRB Coordinates Network, 12573, 1

\bibitem[\protect\citeauthoryear{{Fong} et~al.}{{Fong} et~al.}{2011a}]{Fong11b}
{Fong}, W., et~al. 2011a, \apj, 730, 26

\bibitem[\protect\citeauthoryear{{Fong}, {Berger}, \& {Fox}}{{Fong}
  et~al.}{2010}]{Fong10}
{Fong}, W., {Berger}, E.,  \& {Fox}, D.~B. 2010, \apj, 708, 9

\bibitem[\protect\citeauthoryear{{Fong} et~al.}{{Fong} et~al.}{2011b}]{Fong11}
{Fong}, W., {Sanders}, N., {Milisavljevic}, D.,  \& {Berger}, E. 2011b, GRB
  Coordinates Network, 12566, 1

\bibitem[\protect\citeauthoryear{{Fong},{Zauderer},{Berger}}{{Fong} et~al.}{2011c}]{Fong11c}
{Fong}, W., {Zauderer}, B.~A, {Berger}, E. 2011c, GRB
  Coordinates Network, 12571, 1


\bibitem[\protect\citeauthoryear{{Fong} et~al.}{{Fong} et~al.}{2012}]{Fong12}
{Fong}, W.-f., et~al. 2012, arXiv:1204.5475

\bibitem[\protect\citeauthoryear{{Fox} et~al.}{{Fox} et~al.}{2005}]{Fox05}
{Fox}, D.~B., et~al. 2005, \nat, 437, 845

\bibitem[\protect\citeauthoryear{{Frail} et~al.}{{Frail}
  et~al.}{2001}]{Frail01}
{Frail}, D.~A., et~al. 2001, \apjl, 562, L55

\bibitem[\protect\citeauthoryear{{Fryer}, {Woosley}, \& {Hartmann}}{{Fryer}
  et~al.}{1999}]{Fryer99}
{Fryer}, C.~L., {Woosley}, S.~E.,  \& {Hartmann}, D.~H. 1999, \apj, 526, 152

\bibitem[\protect\citeauthoryear{{Gehrels} et~al.}{{Gehrels}
  et~al.}{2004}]{Gehrels04}
{Gehrels}, N., et~al. 2004, \apj, 611, 1005

\bibitem[\protect\citeauthoryear{{Graham} et~al.}{{Graham}
  et~al.}{2009}]{Graham09}
{Graham}, J.~F., et~al. 2009, \apj, 698, 1620

\bibitem[\protect\citeauthoryear{{Granot} \& {Sari}}{{Granot} \&
  {Sari}}{2002}]{Granot02}
{Granot}, J.,  \& {Sari}, R. 2002, \apj, 568, 820

\bibitem[\protect\citeauthoryear{{Greiner}et~al.}{{Greiner} et~al.}{2008}]{Greiner08}
{Greiner}, J. ,et~al. 2008, \pasp, 120, 405

\bibitem[\protect\citeauthoryear{{Greisen}}{{Greisen}}{2003}]{Greisen03}
{Greisen}, E.~W. ,2003, AIPS, 285, 109

\bibitem[\protect\citeauthoryear{{Grupe} et~al.}{{Grupe} et~al.}{2006}]{Grupe06}
{Grupe}, D., {Burrows}, D.~N., {Patel}, S.~K., {Kouveliotou}, C., {Zhang},
  B., {M{\'e}sz{\'a}ros}, P.,  {Wijers}, R.~A. \& {Gehrels}, N., 2006, \apj, 653, 462

\bibitem[\protect\citeauthoryear{{Hill} et~al.}{{Hill} et~al.}{2004}]{Hill04}
{Hill}, J., et al., 2004, SPIE Conf. Series, 5165, 217

\bibitem[\protect\citeauthoryear{{Hjorth} et~al.}{{Hjorth}
  et~al.}{2005}]{Hjorth05}
{Hjorth}, J., et~al. 2005, \nat, 437, 859

\bibitem[\protect\citeauthoryear{{Hogg} et~al.}{{Hogg} et~al.}{1997}]{Hogg97}
{Hogg}, D.~W., {Pahre}, M.~A., {McCarthy}, J.~K., {Cohen}, J.~G., {Blandford},
  R., {Smail}, I.,  \& {Soifer}, B.~T. 1997, \mnras, 288, 404

\bibitem[\protect\citeauthoryear{{Hook} et~al.}{{Hook} et~al.}{2004}]{Hook04}
{Hook}, I.~M., {J{\o}rgensen}, I., {Allington-Smith}, J.~R., {Davies}, R.~L.,
  {Metcalfe}, N., {Murowinski}, R.~G.,  \& {Crampton}, D. 2004, \pasp, 116, 425

\bibitem[\protect\citeauthoryear{{Ioka}, {Kobayashi}, \& {Zhang}}{{Ioka}
  et~al.}{2005}]{Ioka05}
{Ioka}, K., {Kobayashi}, S.,  \& {Zhang}, B. 2005, \apj, 631, 429

\bibitem[\protect\citeauthoryear{{Kalberla} et~al.}{{Kalberla}
  et~al.}{2005}]{Kalberla05}
{Kalberla}, P.~M.~W., {Burton}, W.~B., {Hartmann}, D., {Arnal}, E.~M.,
  {Bajaja}, E., {Morras}, R.,  \& {P{\"o}ppel}, W.~G.~L. 2005, \aap, 440, 775

\bibitem[\protect\citeauthoryear{{Kaneko} et~al.}{{Kaneko}
  et~al.}{2006}]{Kaneko06}
{Kaneko}, Y., {Preece}, R.~D., {Briggs}, M.~S., {Paciesas}, W.~S., {Meegan},
  C.~A.,  \& {Band}, D.~L. 2006, \apjs, 166, 298

\bibitem[\protect\citeauthoryear{{Kennicutt}}{{Kennicutt}}{1998}]{Kennicutt98}
{Kennicutt}, R.~C., Jr. 1998, \araa, 36, 189

\bibitem[\protect\citeauthoryear{{Lazzati} et~al.}{{Lazzati}
  et~al.}{2011}]{Lazzati11}
{Lazzati}, D., {Blackwell}, C.~H., {Morsony}, B.~J.,  \& {Begelman}, M.~C.
  2011, \mnras, 411, L16

\bibitem[\protect\citeauthoryear{{Leibler} \& {Berger}}{{Leibler} \&
  {Berger}}{2010}]{Leibler10}
{Leibler}, C.~N.,  \& {Berger}, E. 2010, \apj, 725, 1202

\bibitem[\protect\citeauthoryear{{Levan} et~al.}{{Levan}
  et~al.}{2006a}]{Levan06b}
{Levan}, A.~J., et~al. 2006a, \apjl, 648, L9

\bibitem[\protect\citeauthoryear{{Levan} et~al.}{{Levan}
  et~al.}{2006b}]{Levan06}
{Levan}, A.~J., {Wynn}, G.~A., {Chapman}, R., {Davies}, M.~B., {King}, A.~R.,
  {Priddey}, R.~S.,  \& {Tanvir}, N.~R. 2006b, \mnras, 368, L1

\bibitem[\protect\citeauthoryear{{Mangano} et~al.}{{Mangano}
  et~al.}{2011}]{Mangano11}
{Mangano}, V., et~al. 2011, GRB Coordinates Network, 12559, 1

\bibitem[\protect\citeauthoryear{{Mangano} et~al.}{{Mangano}
  et~al.}{2011b}]{Mangano11b}
{Mangano}, V., {Baumgartner}, W.~H., {Krimm}, H.~A.,  {Oates}, S.~R., {Barthelmy}, S.~D., {Burrows}, D.~N.,
{Roming}, P. \& {Gehrels}, N., 2011b, GCN Rep. 363


\bibitem[\protect\citeauthoryear{{Maraston}}{{Maraston}}{2005}]{Maraston05}
{Maraston}, C. 2005, \mnras, 362, 799

\bibitem[\protect\citeauthoryear{{Margutti} et~al.}{{Margutti}
  et~al.}{2011}]{Margutti11}
{Margutti}, R., et~al. 2011, \mnras, 417, 2144

\bibitem[\protect\citeauthoryear{{Margutti} et~al.}{{Margutti}
  et~al.}{2012}]{Margutti12}
{Margutti}, R., et~al. 2012, arXiv:1203.1059

\bibitem[\protect\citeauthoryear{{Melandri} et~al.}{{Melandri}
  et~al.}{2011a}]{Melandri11b}
{Melandri}, A., {Fugazza}, D., {Covino}, S.,  \& {Palazzi}, E. 2011a, GRB
  Coordinates Network, 12570, 1

\bibitem[\protect\citeauthoryear{{Melandri} et~al.}{{Melandri}
  et~al.}{2011b}]{Melandri11}
{Melandri}, A., et~al. 2011b, GRB Coordinates Network, 12565, 1

\bibitem[\protect\citeauthoryear{{Metzger}, {Quataert}, \&
  {Thompson}}{{Metzger} et~al.}{2008}]{Metzger08}
{Metzger}, B.~D., {Quataert}, E.,  \& {Thompson}, T.~A. 2008, \mnras, 385, 1455

\bibitem[\protect\citeauthoryear{{Oates} \& {Mangano}}{{Oates} \&
  {Mangano}}{2011}]{Oates11}
{Oates}, S.~R.,  \& {Mangano}, V. 2011, GRB Coordinates Network, 12569, 1

\bibitem[\protect\citeauthoryear{{Perley} et~al.}{{Perley}
  et~al.}{2009}]{Perley09}
{Perley}, D.~A., et~al. 2009, \apj, 696, 1871

\bibitem[\protect\citeauthoryear{{Perley},{Chandler},{Butler},{Wrobel}}{{Perley}
  et~al.}{2011}]{rperley+11}
{Perley}, R.~A., {Chandler}, C.~J., {Butler}, B.~J., {Wrobel}, J.~M., 2011, \apjl, 739, L1

\bibitem[\protect\citeauthoryear{{Perna} \& {Belczynski}}{{Perna} \&
  {Belczynski}}{2002}]{Perna02}
{Perna}, R.,  \& {Belczynski}, K. 2002, \apj, 570, 252

\bibitem[\protect\citeauthoryear{{Predehl} \& {Schmitt}}{{Predehl} \&
  {Schmitt}}{1995}]{Predehl95}
{Predehl}, P.,  \& {Schmitt}, J.~H.~M.~M. 1995, \aap, 293, 889

\bibitem[\protect\citeauthoryear{{Roming} et~al.}{{Roming}
  et~al.}{2005}]{Roming05}
{Roming}, P.~W.~A., et~al. 2005, \ssr, 120, 95

\bibitem[\protect\citeauthoryear{{Sakamoto} et~al.}{{Sakamoto}
  et~al.}{2011a}]{Sakamoto11}
{Sakamoto}, T., et~al. 2011a, GRB Coordinates Network, 12561, 1

\bibitem[\protect\citeauthoryear{{Sakamoto} et~al.}{{Sakamoto}
  et~al.}{2011b}]{Sakamoto11b}
{Sakamoto}, T., {Troja}, E., {Gehrels}, N., {Norris}, J., {Barthelmy}, S.~D.,
  {Racusin}, J.~L., {Kawai}, N.,  \& {Fruchter}, A. 2011b, GRB Coordinates
  Network, 12580, 1

\bibitem[\protect\citeauthoryear{{Sakamoto} et~al.}{{Sakamoto}
  et~al.}{2012}]{Sakamoto12}
{Sakamoto}, T., et~al. 2012, arXiv:1205.6774

\bibitem[\protect\citeauthoryear{{Schlafly} \& {Finkbeiner}}{{Schlafly} \&
  {Finkbeiner}}{2011}]{Schlafly11}
{Schlafly}, E.~F.,  \& {Finkbeiner}, D.~P. 2011, \apj, 737, 103

\bibitem[\protect\citeauthoryear{{Schmidl} et~al.}{{Schmidl}
  et~al.}{2011}]{Schmidl11}
{Schmidl}, S., {Rossi}, A., {Kann}, D.~A.,  \& {Greiner}, J. 2011, GRB
  Coordinates Network, 12568, 1

\bibitem[\protect\citeauthoryear{{Soderberg} et~al.}{{Soderberg}
  et~al.}{2006}]{Soderberg06}
{Soderberg}, A.~M., et~al. 2006, \apj, 650, 261

\bibitem[\protect\citeauthoryear{{Willmer} et~al.}{{Willmer}
  et~al.}{2006}]{Willmer06}
{Willmer}, C.~N.~A., et~al. 2006, \apj, 647, 853

\bibitem[\protect\citeauthoryear{{Zhang} et~al.}{{Zhang}
  et~al.}{2006}]{Zhang06}
{Zhang}, B., {Fan}, Y.~Z., {Dyks}, J., {Kobayashi}, S., {M{\'e}sz{\'a}ros}, P.,
  {Burrows}, D.~N., {Nousek}, J.~A.,  \& {Gehrels}, N. 2006, \apj, 642, 354

\bibitem[\protect\citeauthoryear{{Zhao} et~al.}{{Zhao} et~al.}{2011}]{Zhao11}
{Zhao}, X.-H., {Xu}, D., {Mao}, J.-R.,  \& {Bai}, J.-M. 2011, GRB Coordinates
  Network, 12560, 1

\end{thebibliography}

\end{document}